\documentclass[a4paper,11pt]{article}

\usepackage{latexsym}
\usepackage{amssymb}
\usepackage{amsfonts}
\usepackage{amsmath}

\renewcommand{\theequation}{\thesection.\arabic{equation}}
\newcommand{\bol}[1]{\boldsymbol {#1}}
\newcommand{\sss}[1]{\scriptscriptstyle {#1}}

\def\d{\partial}
\def\f{\phi}

\begin{document}

\author{Nicola Grillo\thanks{\texttt{grillo@physik.unizh.ch}} \\  \\
{\textit{Institut f\"ur Theoretische Physik, Universit\"at Z\"urich}} \\
{\textit{Winterthurerstrasse 190, CH-8057 Z\"urich, Switzerland} }}

\title{Scalar Matter Coupled to Quantum Gravity in the Causal Approach: Finite One-Loop Calculations and Perturbative Gauge Invariance}

\maketitle

\begin{abstract}
\noindent
Quantum gravity coupled to scalar massive matter  fields is investigated in the framework of causal perturbation theory.
One-loop calculations include matter loop graviton self-energy and matter self-energy and yield ultraviolet finite and 
cutoff-free expressions. Perturbative gauge invariance to second order  implies the usual Slavnov--Ward 
identities for the graviton self-energy in the loop graph sector and generates the correct quartic graviton-matter interaction in 
the tree graph sector. The mass zero case is also discussed.

\vskip 7mm
{\bf Preprint:} ZU-TH 40/1999
\end{abstract}
\newpage

\tableofcontents

\newpage

\section{Introduction}
\setcounter{equation}{0}

In this paper we follow the quantum field theoretical approach to gravitational interactions coupled to scalar matter 
fields (see the introduction to this subject in~\cite{feynman} and reference therein). This approach allows a quantization of 
the involved  fields, matter and graviton fields, and a Lorentz covariant perturbative expansion of the scattering matrix $S$.

Calculations of matter loop diagrams in this conventional  framework led to non-renormalizable ultraviolet (UV) divergences~\cite{fey1}. 
These were later confirmed by means of dimensional regularization and background field method, both in the massive~\cite{cap3} 
and in the massless~\cite{hooft},~\cite{velt} case.

The counterterms needed to cancel the divergences are not of the type present in the original Lagrangian density. According to these 
findings, quantum gravity (QG) coupled to matter fields  does not fulfil the criterion of perturbative renormalizability~\cite{des8}.

We show how it is possible to overcome these discouraging outcomes by applying an improved perturbation scheme which has as central 
objects the time-ordered products and as constructing principle causality. The $S$-matrix is constructed inductively as a sum of 
smeared operator-valued $n$-point distributions avoiding UV divergences. 

This idea goes back to St\"uckelberg, Bogoliubov and Shirkov and the program (causal perturbation theory) was carried out successfully 
by Epstein and Glaser~\cite{eg} for scalar field theories and subsequently applied to QED by Scharf~\cite{scha}, to non-Abelian gauge 
theories by D\"utsch {\it et al.}~\cite{ym2} and to quantum gravity~\cite{scho1} in the last years.

In addition, QG has considerable gauge properties~\cite{well1}, which are formulated by means of a `gauge charge' that generates 
infinitesimal gauge variations of the fundamental free quantum fields, Sec~\ref{sec:23}.

The present work focuses mainly on three aspects of QG coupled to massive matter fields. Brief remarks for the massless case are given 
for these aspects.
The first aspect  is the UV finiteness of loop graphs, which include the lowest order  massive and massless scalar matter loop
corrections to the graviton propagator, Sec.~\ref{sec:3},  and the matter self-energy, Sec.~\ref{sec:4}. The result are UV finite and 
cutoff-free due to techniques of  causal perturbation theory.

The second aspect consists in the investigation of the gauge properties of the graviton self-energy, Sec.~\ref{sec:24}.
Gauge invariance of the $S$-matrix implies some identities between the C-number parts of the $n$-point distributions which yield
the gravitational Slavnov--Ward identities (SWI)~\cite{capSW}, Sec.~\ref{sec:33}.

The third aspect of this work is also connected with gauge invariance: perturbative gauge invariance to second order in the tree
graph sector requires the introduction, at a purely quantum level, of a quartic matter-graviton interaction exactly as prescribed
by the expansion of the classical matter-gravity Lagrangian, Sec.~\ref{sec:5}.

The quantization of the graviton field, the identification of the physical subspace and the proof of $S$-matrix unitarity are 
investigated in~\cite{gri3}, which provides also the conventions and the notations used here. Calculations involving graviton 
self-couplings are not considered here, see~\cite{gri4}.

The causal scheme applied to quantum gravity coupled to photon fields leads also to very satisfactory results with regard to
the UV finiteness in loop calculations and to the gauge invariance~\cite{gri5}.

We use the unit convention: $\hbar=c=1$, Greek indices $\alpha,\beta,\ldots$ run from $0$ to $3$, whereas Latin 
indices $i,j,\ldots$ run from $1$ to $3$.

\section{Quantized Matter--Gravity System and Perturbative Gauge Invariance}\label{sec:2}
\setcounter{equation}{0}

\subsection{Inductive Construction of Two-Point Distributions in the\\ \boldmath{$S$}-Matrix Expansion}\label{sec:21}

In causal perturbation theory~\cite{scha},~\cite{aste}, the ansatz for the $S$-matrix as a power 
series in the coupling constant is central, namely $S$ is considered as a sum of smeared operator-valued distributions
\begin{equation}
S(g)={\mathbf 1}+\sum_{n=1}^{\infty}\frac{1}{n!} \int\!\! d^{4}x_{1}
\ldots d^{4}x_{n}\, T_{n}(x_{1},\ldots, x_{n})\,g(x_{1})\cdot\ldots\cdot g(x_{n})\,  ,
\label{eq:d1}
\end{equation}
where the  Schwartz test function $g\in\mathcal{S}(\mathbb{R}^{4})$ switches the interaction and provides a natural 
infrared cutoff. The $S$-matrix maps the asymptotically incoming free fields on the outgoing ones and it is possible
to express the $T_{n}$'s  by means of free fields without introducing interacting quantum fields.

The $n$-point distribution $T_{n}$ is a  well-defined  `renormalized' time-ordered product expressed 
in terms of Wick monomials of free fields $:\!{\mathcal O}(x_{1},\ldots,x_{n})$:
\begin{equation}
T_{n}(x_{1},\ldots, x_{n})=\sum_{\sss { \mathcal{ O}}} :\!{\mathcal O}(x_{1},\ldots,x_{n})\!:\,
t_{n}^{\sss{ \mathcal O}}(x_{1}-x_{n},\ldots,x_{n-1}-x_{n})\,.
\label{eq:d2}
\end{equation}
The $t_{n}$' are C-number distributions. $T_{n}$ is constructed inductively from the first order $T_{1}(x)$,
which describes the interaction among the quantum fields,  and from 
the lower orders $T_{j}$, $j=2,\ldots,n-1$ by means of Poincar\'e covariance and causality. The latter leads directly 
to a UV finite and cutoff-free distribution $T_{n}$. 

For the purpose of this paper, we outline briefly the main steps in the construction of $T_{2}(x,y)$ from 
a given  first order interaction. Following the inductive scheme, we first calculate the causal operator-valued distribution
\begin{equation}
D_{2}(x,y):=R'_{2}(x,y)-A'_{2}(x,y)=\big[T_{1}(x),T_{1}(y)\big]\,.
\label{eq:d4}
\end{equation}
In order to obtain $D_{2}(x,y)$, one has to carry out all possible contractions between the normally ordered $T_{1}$ 
using Wick's lemma, so that $D_{2}(x,y)$ has the following structure
\begin{equation}
D_{2}(x,y)=\sum_{\sss { \mathcal{ O}}} :\!{\mathcal O}(x,y)\!:\,d_{2}^{\sss{ \mathcal O}}(x-y)\,.
\label{eq:d5}
\end{equation}
$d_{2}^{\sss {\mathcal O}}(x-y) $ is a numerical distribution that depends only on the relative coordinate $x-y$,
because of translation invariance. 

$D_{2}(x,y)$ contains tree (one contraction), loop (two contractions) and vacuum graph (three contractions)
contributions. Due to the presence of normal ordering, tadpole diagrams do not show up. In this paper we do not consider
vacuum graphs. $D_{2}(x,y)$ is causal,
\emph{i.e.} $\hbox{supp}(d_{2}^{\sss{\mathcal O}} (z))\subseteq \overline{ V^{\sss +}(z)}\cup \overline{ V^{\sss -}(z)}$, 
with $z:=x-y$.

In order to obtain $T_{2}(x,y)$, we have to split $D_{2}(x,y)$ into a retarded part, $R_{2}(x,y)$, and
an advanced part, $A_{2}(x,y)$, with respect to the coincident point $z=0$, so that 
$\hbox{supp}(R_{2}(z))\subseteq \overline{ V^{\sss +}(z)}$ and 
$\hbox{supp}(A_{2}(z))\subseteq \overline{ V^{\sss -}(z)}$.
This splitting, or `time-ordering', has to be carried out in the distributional sense so that
the retarded and advanced part are well-defined and UV finite.

The splitting affects only the numerical distribution $d_{2}^{\sss {\mathcal O}}(x-y)$ and must be accomplished
according to the correct `singular order' $\omega(d_{2}^{\sss {\mathcal O}})$ which describes roughly speaking the 
behaviour of  $d_{2}^{\sss{\mathcal O}}(x-y)$ near $x-y=0$, or that of $\hat{d}_{2}^{\sss {\mathcal O}}(p)$ in the limit 
$p\to\infty$. If $\omega < 0$, then the splitting is trivial and agrees with the standard time-ordering. 
If $\omega\ge 0$,
then the splitting is non-trivial and non-unique: 
\begin{equation}
d_{2}^{\sss{\mathcal O}}(x-y)\longrightarrow r_{2}^{\sss {\mathcal O}}(x-y)+
\sum_{|a|=0}^{\omega(d_{2}^{\sss {\mathcal O}})}\,C_{a,\sss{\mathcal O}}\, D^{a}\, \delta^{\sss (4)}(x-y)\,.
\label{eq:d6}
\end{equation}
A  retarded part $r_{2}^{\sss{\mathcal O}}(x-y)$ of $d_{2}^{\sss{\mathcal O}}(x-y)$ is usually obtained in momentum space  
by means of a dispersion-like integral, see Eq.~(\ref{eq:d48}).

Eq.~(\ref{eq:d6}) contains a local ambiguity in the normalization: the $C_{a,\sss{\mathcal O}}$'s are undetermined finite 
normalization constants, which multiply terms with point support $D^{a}\delta^{\sss (4)}(x-y)$ ($D^{a}$ is a partial 
differential operator). This freedom in the normalization has to be restricted by  physical conditions.

Finally, $T_{2}$ is obtained  by subtracting $R'_{2}(x,y)$ from $R_{2}(x,y)$ and the whole local normalization coming
from~(\ref{eq:d6}) is called  $N_{2}(x,y)$.

\subsection{Quantized Matter-Gravity Interaction}\label{sec:22}

We consider the coupling between the quantized symmetric tensor field $h^{\mu\nu}(x)$, the graviton, and 
the quantized scalar field $\f(x)$, the matter field, in the background of a Minkowski space-time.

The free scalar field of mass $m$ satisfies the Klein--Gordon wave equation
\begin{equation}
(\Box +m^2)\f (x)=0\,,
\label{eq:d9}
\end{equation}
which follows from the free matter Lagrangian density 
$\mathcal{L}_{\sss M}^{\sss (0)}=\frac{1}{2}\f_{,\nu}\f^{,\nu}-\frac{m^2}{2}\f^2$. The matter energy-momentum tensor reads 
\begin{equation}
T_{\sss M}^{\mu\nu}:=\f^{,\mu}\f^{,\nu}-\eta^{\mu\nu}\,\mathcal{L}_{\sss M}^{\sss (0)}
    =\f^{,\mu}\f^{,\nu}-\frac{1}{2}\eta^{\mu\nu}\,\f_{,\rho}\f^{,\rho}+\eta^{\mu\nu}\,\frac{m^2}{2}\f^2\,,
\label{eq:d10}
\end{equation}
where $\eta^{\mu\nu}=\mathrm{diag}(+1,-1,-1,-1)$ and it fulfils $T_{{\sss M}\, ,\nu}^{\mu\nu}=0$. Quantization of the scalar field
is accomplished through
\begin{equation}
\big[\f(x),\f(y)\big]=-i\,D_{m}(x-y)\,,
\label{eq:d11}
\end{equation}
where
\begin{equation}
D_{m}(x)=D^{\sss (+)}_{m}(x)+D^{\sss (-)}_{m}(x)=\frac{i}{(2\pi)^3}\int \!\!d^{4}p\, \delta (p^2 -m^2)\,
     \mathrm{sgn}(p^{0})\, e^{-i\,p\cdot x} 
\label{eq:d12}
\end{equation}
is the causal Jordan--Pauli distribution of mass $m$.

The free graviton field satisfies the wave equation
\begin{equation}
\Box\,h^{\mu\nu}(x)=0
\label{eq:d13}
\end{equation}
and is quantized (see~\cite{gri3}) according to 
\begin{equation}
\big[ h^{\mu\nu}(x),h^{\alpha\beta}(y)\big]=-i\,b^{\mu\nu\alpha\beta}\,D_{0}(x-y)\,,
\label{eq:d14}
\end{equation}
where the $b$-tensor is constructed from the Minkowski metric
\begin{equation}
b^{\mu\nu\alpha\beta}:=\frac{1}{2}\left(\eta^{\mu\alpha}\eta^{\nu\beta}+\eta^{\mu\beta}\eta^{\nu\alpha}-
\eta^{\mu\nu}\eta^{\alpha\beta}\right)
\label{eq:d14.1}
\end{equation}
and $D_{0}(x)$ is the  Jordan--Pauli distribution of Eq.~(\ref{eq:d12}) with $m=0$.

The graviton field interacts with the conserved energy-momentum tensor of the matter fields. The first order matter coupling
is chosen to be
\begin{equation}
T_{1}^{\sss M}(x)=i\,\frac{\kappa}{2}:\!h^{\alpha\beta}(x)\,b_{\alpha\beta\mu\nu}\,T_{\sss M}^{\mu\nu}(x)\!:=
i\,\frac{\kappa}{2}\,\big\{:\!h^{\alpha\beta}\f_{,\alpha}\f_{,\beta}\!:-\frac{m^2}{2}:\!h\f\f\!:\big\}\,,
\label{eq:d15}
\end{equation}
where $\kappa$ is the coupling constant (see below for its relation to Newton's constant).

To simplify the notation, the trace of the graviton field is written as $h=h^{\gamma}_{\  \gamma}$ and all Lorentz 
indices of the fields are written as superscripts whereas the derivatives acting on the fields are written as 
subscripts. All indices occurring twice are contracted by the Minkowski metric $\eta^{\mu\nu}$. We skip the 
space-time dependence if the meaning is clear.

The presence of the $b$-tensor in Eq.~(\ref{eq:d15}) is a consequence of the choice of the graviton variable (see Sec.~\ref{sec:23}).

\subsection{Perturbative Gauge Invariance}\label{sec:23}

The classical gauge properties of $h^{\mu\nu}(x)$ (which are related to the general covariance of the metric $g_{\mu\nu}(x)$
under coordinate transformations~\cite{wyss},~\cite{wald}) are formulated at  the  quantum level by the gauge 
charge~\cite{scho1},~\cite{well1}
\begin{equation}
Q:=\int\limits_{x^0 =const}\!\! d^{3}x\,h^{\mu\nu}(x)_{,\nu} {\stackrel{\longleftrightarrow}{ \partial_{x}^{0} }
 u_{\mu}(x)} \, .
\label{eq:d16}
\end{equation}
For the construction of the physical subspace  and in order to prove unitarity of the $S$-matrix on the 
physical subspace~\cite{gri3}, the ghost field $u^{\mu}(x)$, together with the anti-ghost field 
$\tilde{u}^{\nu}(x)$, have to be quantized as free fermionic vector fields:
\begin{equation}
\Box u^{\mu}(x)=0\,,\quad \Box \tilde{u}^{\nu}(x)=0\,,\quad 
\big\{u^{\mu}(x),\tilde{u}^{\nu}(y)\big\}=i\, \eta^{\mu\nu}\, D_{0}(x-y) \, ,
\label{eq:d17}
\end{equation}
whereas all other anti-commutators vanish. The gauge charge generates the following infinitesimal operator gauge variations
\begin{gather}
d_Q h^{\mu\nu}(x) : =\big[ Q, h^{\mu\nu}(x)\big]=-i\, b^{\mu\nu\rho\sigma} u^{\rho}(x)_{,\sigma}\,,\nonumber\\
d_Q u^{\alpha}(x) := \big\{ Q, u^{\alpha}(x)\big\}=0\,,\quad
d_Q \tilde{u}^{\alpha}(x) := \big\{ Q, \tilde{u}^{\alpha}(x)\big\}=i\, h^{\alpha\beta}(x)_{,\beta}\,.
\label{eq:d18}
\end{gather}
Gauge invariance of the $S$-matrix means formally
\begin{equation}
\lim_{g\to 1} d_Q S(g)=0\,.
\label{eq:d19}
\end{equation}
This condition can be reformulated in terms on the $n$-point distributions $T_{n}$: using
Eq.~(\ref{eq:d1}), we see that the condition of perturbative gauge invariance to $n$-th order
\begin{equation}
d_Q T_{n}(x_{1},\ldots,x_{n})=\text{sum of divergences}\,,
\label{eq:d20}
\end{equation}
implies Eq.~(\ref{eq:d19}), because divergences do not contribute in the adiabatic limit $g\to 1$ due to 
partial integration and Gauss' theorem.

Using  a simplified notation which keeps track of the field type only, perturbative invariance to first order in pure QG~\cite{scho1}, 
namely for a coupling of the form $T_{1}^{h}\sim :\!hhh\!:$ without matter fields, requires the introduction of the ghost 
coupling  $T_{1}^{u}\sim :\!\tilde{u}h u\!:$ , so that
$d_{Q}(T_{1}^{h}+T_{1}^{u})(x)=\d_{\sigma}^{x}\,T_{1/1}^{\sigma\: h+u}(x)$. Here, 
$T_{1/1}^{\sigma\: h+u}\sim :\!uhh\!:+:\!\tilde{u}uu\!:$
is the so-called $Q$-vertex.

Using Eq.~(\ref{eq:d18}), the first order matter coupling~(\ref{eq:d15}) is gauge invariant:
\begin{equation}
\begin{split}
d_{Q} T_{1}^{\sss M}(x)&=\,\frac{\kappa}{2}\,b^{\alpha\beta\rho\sigma}:\!u^{\rho}(x)_{,\sigma}\,b_{\alpha\beta\mu\nu}\,
T_{\sss M}^{\mu\nu}(x)\!:=\frac{\kappa}{2}:\!u^{\rho}(x)_{,\sigma}\,T_{\sss M}^{\rho\sigma}(x)\!:\\
&=\,\d_{\sigma}^{x}\Big(\frac{\kappa}{2}\,\big\{:\!u^{\rho}\f_{,\rho}\f_{,\sigma}\!:-\frac{1}{2}:\!u^{\sigma}\f_{,\rho}\f_{,\rho}\!:
+\frac{m^2}{2}:\!u^{\sigma}\f\f\!:\big\}\Big)\\
&=:\d_{\sigma}^{x}\,T_{1/1}^{\sigma\; \sss M}(x)\,,
\label{eq:d21}
\end{split}
\end{equation}
because of $T_{{\sss M}\: ,\sigma}^{\rho\sigma}=0$.  $T_{1/1}^{\sigma\; \sss M}(x)$ is the matter $Q$-vertex.
The concept of $Q$-vertex allows us to formulate in a precise way the condition~(\ref{eq:d20}) of perturbative gauge invariance to 
the $n$-th order:
\begin{equation}
d_{Q}T_{n}(x_{1},\ldots ,x_{n})=\sum_{l=1}^{n}\frac{\partial}{\partial x^{\nu}_{l}}\, 
T_{n/l}^{\nu}(x_{1},\ldots , x_{l},\ldots ,x_{n}) \, ,
\label{eq:d22}
\end{equation}
where  $T_{n/l}^{\nu}$ is the `renormalized' time-ordered product, obtained according to the inductive causal scheme, 
with one $Q$-vertex at $x_{l}$, while all other $n-1$ vertices are ordinary $T_{1}$-vertices.

The procedure outlined here corresponds to the expansion of the Hilbert--Einstein and matter Lagrangian 
density~\cite{fey1},~\cite{cap3},~\cite{velt}
\begin{equation}
\mathcal{L}_{\sss EH}+\mathcal{L}_{\sss M}=\frac{-2}{\kappa^2}\,\sqrt{-g}\,g^{\mu\nu}\,R_{\mu\nu} +\frac{1}{2}\sqrt{-g}\,
\big(g^{\mu\nu}\f_{;\mu}\f_{;\nu}-m^2\,\f^2\big)\,,
\label{eq:d23}
\end{equation}
($\kappa^2=32\pi\,G$) written in terms of the Goldberg variable $\tilde{g}^{\mu\nu}$, in powers of the coupling
constant $\kappa$ according to the metric decomposition
\begin{equation}
\tilde{g}^{\mu\nu}:=\sqrt{-g}\,g^{\mu\nu}=\eta^{\mu\nu}+\kappa\,h^{\mu\nu}\,,
\label{eq:d24}
\end{equation}
which defines the graviton field $h^{\mu\nu}$ in the Minkowski background. Then one obtains
\begin{equation}
\mathcal{L}_{\sss EH}+\mathcal{L}_{\sss M}=\sum_{j=0}^{\infty}\,\kappa^{j}\,\big(\mathcal{L}^{\sss (j)}_{\sss EH}+
          \mathcal{L}^{\sss (j)}_{\sss M}\big)\,.
\label{eq:d25}
\end{equation}
From $\mathcal{L}^{\sss (0)}_{\sss EH}$, choosing the Hilbert gauge $h^{\mu\nu}_{\:,\nu}=0$, one obtains Eq.~(\ref{eq:d13}) and the
presence of the $b$-tensor is made clear~\cite{gri3}. 

The first order graviton coupling $T_{1}^{h}(x)\sim :\!hhh\!:$, corresponds then to the normally ordered product 
of $i\kappa\,\mathcal{L}^{\sss (1)}_{\sss EH}$ (see~\cite{well1} for a derivation based merely on the principle of perturbative
operator gauge invariance) and 
$\mathcal{L}^{\sss (2)}_{\sss EH}\sim h h h h$ represents
the quartic graviton coupling required by perturbative gauge invariance to second order in the tree graph sector~\cite{scho1}.

The expansion of the matter Lagrangian density reads
\begin{multline}
\mathcal{L}_{\sss M}=\underbrace{\frac{1}{2}\,\big(\eta^{\mu\nu}\f_{,\mu}\f_{,\nu}-m^2\,\f^2 \big)}_
{=\mathcal{L}_{\sss M}^{\sss (0)} } 
+\underbrace{\frac{\kappa}{2}\, h^{\mu\nu}\,\big(\f_{,\mu}\f_{,\nu}-\frac{m^2}{2}\,\eta_{\mu\nu}\,\f^2 \big)}_
{=\kappa\, \mathcal{L}_{\sss M}^{\sss (1)} } \\
+\underbrace{ \frac{m^2 \kappa^2}{8}\,\big( h^{\alpha\beta}h^{\alpha\beta}\f \f
   -\frac{1}{2}\, h h\f \f \big)}_{=\kappa^2\,\mathcal{L}_{\sss M}^{\sss (2)} }+ O(\kappa^3)\, .
\label{eq:d26}
\end{multline}
From the first term one obtains~(\ref{eq:d9}) and the matter coupling of Eq.~(\ref{eq:d15}) corresponds to 
$i\,\kappa\,:\!\mathcal{L}_{\sss M}^{\sss (1)}\!:$. A quantized quartic interaction $\sim:\!h h \f \f \!:$ 
which agrees with $\mathcal{L}_{\sss M}^{\sss (2)}$ will be necessary for reason of gauge invariance, 
see Sec.~\ref{sec:5}.

\subsection{Identities for the Two-Point Functions from Perturbative Gauge Invariance to Second Order}\label{sec:24}

From the structure of $T_{1}^{\sss M}$ it is evident that the two-point distribution describing loop graphs has the form
(up to non-contributing divergences for the matter self-energy, see Sec.~\ref{sec:41}):
\begin{equation}
T_{2}(x,y)^{\text{loops}}=:\!h^{\alpha\beta}(x)h^{\mu\nu}(y)\!:\,i\,\Pi(x-y)_{\alpha\beta\mu\nu}+
:\!\f(x)\f(y)\!:\,i\,\Sigma(x-y)\,.
\label{eq:d27}
\end{equation}
Here, the first term represents the matter loop graviton self-energy and the second term the scalar matter  self-energy.
The C-number distributions $\Pi(x-y)_{\alpha\beta\mu\nu}$ and $\Sigma(x-y)$ will be explicitly calculated in Sec.~\ref{sec:33} and 
in Sec.~\ref{sec:42}, respectively.

Perturbative gauge invariance to second order, namely Eq.~(\ref{eq:d22}) with $n=2$, allows us to derive a set of identities
for these numerical distributions by comparing distributions attached to the same  operators  on both sides
of Eq.~(\ref{eq:d22}),~\cite{du4}. 

We compute  $d_{Q}T_{2}(x,y)^{\text{loops}}$ by means of~(\ref{eq:d18}) and isolate the contributions with external
operator of the type  $:\!u(x)h(y)\!:$. We obtain
\begin{equation}
d_{Q}T_{2}(x,y)^{\text{loops}}\big|_{:u(x)h(y):}=:\!u^{\rho}(x)_{,\sigma}h^{\mu\nu}(y)\!:\,\big(b^{\alpha\beta\rho\sigma}\,
\Pi(x-y)_{\alpha\beta\mu\nu}\big)\,.
\label{eq:d28}
\end{equation}
On the other side, $T_{2/1}^{\sigma}(x,y)$ has to be constructed with one $Q$-vertex at $x$ and one `normal' vertex at $y$.
From the structure of both interaction terms, it follows that the loop contributions coming from $T_{2/1}^{\sigma}(x,y)$ can only be 
of the form
\begin{equation}
T_{2/1}^{\sigma}(x,y)=:\!u^{\rho}(x)h^{\mu\nu}(y)\!:\,t_{uh}^{\sigma}(x-y)_{\rho\mu\nu}+
:\!u^{\sigma}(x)h^{\mu\nu}(y)\!:\,t_{uh}(x-y)_{\mu\nu}\,,
\label{eq:d29}
\end{equation}
by performing two matter field contractions. The second term $T_{2/2}^{\sigma}(x,y)$ does not contain terms with Wick monomials of 
the type $:\!u(x)h(y)\!:$. Applying $\d_{\sigma}^{x}$ to the expression above we find
\begin{equation}
\begin{split}
\d_{\sigma}^{x}T_{2/1}^{\sigma}(x,y)=&+:\!u^{\rho}(x)_{,\sigma}h^{\mu\nu}(y)\!:\,
\big\{t_{uh}^{\sigma}(x-y)_{\rho\mu\nu}+\eta_{\rho}^{\; \sigma}\,t_{uh}(x-y)_{\mu\nu}\big\}+\\
&+:\!u^{\rho}(x)h^{\mu\nu}(y)\!:\,\d_{\sigma}^{x}\,
\big\{ t_{uh}^{\sigma}(x-y)_{\rho\mu\nu}+\eta_{\rho}^{\; \sigma}\,
t_{uh}(x-y)_{\mu\nu}\big\}\,.
\label{eq:d30}
\end{split}
\end{equation}

We compare the C-number distributions in~(\ref{eq:d28}) and in~(\ref{eq:d30}) attached to the external operators 
\begin{equation}
:\!u^{\rho}(x)_{,\sigma}h^{\mu\nu}(y)\!:\quad\text{and}\quad :\!u^{\rho}(x)h^{\mu\nu}(y)\!:\,.
\label{eq:d31}
\end{equation}
Therefore, we obtain two  identities
\begin{equation}
\begin{split}
b^{\rho\sigma\alpha\beta}\,\Pi(x-y)_{\alpha\beta\mu\nu}&=\big\{t_{uh}^{\sigma}(x-y)^{\rho}_{\ \mu\nu}
+\eta^{\rho\sigma}\,t_{uh}(x-y)_{\mu\nu}\big\}\,,\\
0&=\d_{\sigma}^{x}\,\big\{t_{uh}^{\sigma}(x-y)^{\rho}_{\ \mu\nu} 
+\eta^{\rho\sigma}\,t_{uh}(x-y)_{\mu\nu}\big\}\,.
\label{eq:d32}
\end{split}
\end{equation}
By applying $\d_{\sigma}^{x}$ to the first identity and inserting the second one, we obtain
\begin{equation}
b^{\alpha\beta\rho\sigma}\,\d_{\sigma}^{x}\,\Pi(x-y)_{\alpha\beta\mu\nu}=0\,.
\label{eq:d33}
\end{equation}
This identity for the matter loop graviton self-energy tensor has been explicitly checked and 
implies the gravitational Slavnov--Ward  identities  for the two-point connected Green function (see Sec.~\ref{sec:33}).

Gauge invariance to second order in the tree graph sector is much more involved and requires also the full analysis of the 
matter-graviton interaction, see Sec.~\ref{sec:5}.

\section{Matter Loop Graviton Self-Energy}\label{sec:3}
\setcounter{equation}{0}

\subsection{Causal \boldmath{$D_{2}(x,y)$}-Distribution}\label{sec:31}

In order to construct $D_{2}(x,y)$, according to Sec.~\ref{sec:21}, we first need the contractions between field operators. 
From~(\ref{eq:d11}) and~(\ref{eq:d14}), we  derive them:
\begin{gather}
C\big\{\f(x) \f(y)\big\}:=\big[\f(x)^{\sss (-)},\f(y)^{\sss (+)} \big]= -i\,D^{\sss (+)}_{m}(x-y)\,,\nonumber\\
C\big\{ h^{\alpha\beta}(x)\ h^{\mu\nu}(y) \big\}:=\big[h^{\alpha\beta}(x)^{\sss (-)},h^{\mu\nu}(y)^{\sss (+)}\big]=-i\,
b^{\alpha\beta\mu\nu}\, D_{0}^{\sss (+)}(x-y)\,,
\label{eq:d34}
\end{gather}
where $(\pm)$ refers to the positive/negative frequency part of the corresponding quantity.

The $A'_{2}(x,y)^{g\sss SE}$ distribution for the graviton self-energy  by a matter loop is obtained by 
performing two matter field contractions in $-T_{1}^{\sss M}(x)T_{1}^{\sss M}(y)$. Using~(\ref{eq:d34}) and with 
$\d_{\alpha}^{x}=-\d_{\alpha}^{y}$, we find
\begin{equation}
\begin{split}
A'_{2}(x,y)^{g\sss SE}=&:\!h^{\alpha\beta}(x)h^{\mu\nu}(y)\!:\,\frac{\kappa^2}{4}\,
 \Big[-\d_{\alpha}^{x}\d_{\mu}^{x}D_{m}^{\sss (+)}\cdot\d_{\beta}^{x}\d_{\nu}^{x}D_{m}^{\sss (+)}+\\
      &\qquad - \d_{\alpha}^{x}\d_{\nu}^{x}D_{m}^{\sss (+)}\cdot\d_{\beta}^{x}\d_{\mu}^{x}D_{m}^{\sss (+)}+
   m^2\,\eta_{\mu\nu}\,\d_{\alpha}^{x}D_{m}^{\sss (+)}\cdot\d_{\beta}^{x}D_{m}^{\sss (+)}+\\
 &\qquad + m^2\,\eta_{\alpha\beta}\,\d_{\mu}^{x}D_{m}^{\sss (+)}\cdot\d_{\nu}^{x}D_{m}^{\sss (+)}
-\frac{m^4}{2}\,\eta_{\alpha\beta}\eta_{\mu\nu}\,D_{m}^{\sss (+)}\cdot D_{m}^{\sss (+)}\Big](x-y)\,.    
\raisetag{15mm}\label{eq:d35}
\end{split}
\end{equation}
We introduce the functions
\begin{gather}
D^{\sss (\pm),m}_{\cdot|\cdot}(x):=D_{m}^{\sss (\pm)}(x)\cdot D_{m}^{\sss (\pm)}(x)\,,\qquad
D^{\sss (\pm),m}_{\alpha |\beta}(x):=\partial_{\alpha}^{x} D_{m}^{\sss(\pm)}(x)\cdot\partial_{\beta}^{x} 
                      D_{m}^{\sss (\pm)}(x)\,,\nonumber\\
D_{\alpha\beta |\mu\nu}^{\sss (\pm),m}(x):=\partial_{\alpha}^{x}\partial_{\beta}^{x} D_{m}^{\sss (\pm)}(x)
               \cdot \partial_{\mu}^{x}\partial_{\nu}^{x} D_{m}^{\sss (\pm)}(x)\,,
\label{eq:d36}
\end{gather}
so that we have
\begin{gather}
A'_{2}(x,y)^{g\sss SE}=:\!h^{\alpha\beta}(x)h^{\mu\nu}(y)\!:\,a'_{2}(x-y)_{\alpha\beta\mu\nu}^{g\sss SE}\,,\nonumber\\
\begin{split}a'_{2}(x-y)_{\alpha\beta\mu\nu}^{g\sss SE}=\frac{\kappa^2}{4}\,
\Big[& -D_{\alpha\mu |\beta\nu}^{\sss (\pm),m}-D_{\alpha\nu |\beta\mu}^{\sss (\pm),m}
    +m^2\,\eta_{\mu\nu}\,D^{\sss (\pm),m}_{\alpha |\beta}+\\
  & +  m^2\,\eta_{\alpha\beta}\,D^{\sss (\pm),m}_{\mu|\nu}
-\frac{m^4}{2}\,\eta_{\alpha\beta}\eta_{\mu\nu}\,D^{\sss (\pm),m}_{\cdot|\cdot}\Big](x-y)\,.
\label{eq:d37}
\end{split}
\end{gather}
Products of Jordan--Pauli distributions are evaluated in momentum space with Eq.~(\ref{eq:d12}), because there  products go over into 
convolutions between the Fourier transforms. For example, $D^{\sss (\pm),m}_{\cdot|\cdot}(x)$ becomes 
\begin{equation}
\hat{D}_{\cdot|\cdot}^{\sss (\pm),m}(p)=\frac{-1}{(2\pi)^4}\int\!\!d^4k\,\delta\big((p-k)^2 -m^2\big)\,\Theta\big(\pm(p^0-k^0)\big)\,
\delta(k^2 -m^2)\,\Theta(\pm k^0)\,.
\label{eq:d38}
\end{equation}
Therefore, the basic integrals that remain to be computed are of the form
\begin{equation}
\begin{split}
I_{m}^{\sss (\pm)}(p)_{- / \alpha/\alpha\beta/\alpha\beta\mu/\alpha\beta\mu\nu}:=\int\!\!& d^{4}k
\,\delta\big( (p-k)^2 -m^2 \big)\,
\Theta\big(\pm (p^{0}-k^{0})\big)\,\delta(k^{2} -m^2)\\
&\times\Theta(\pm k^{0})\,\big[1,k_{\alpha},k_{\alpha}k_{\beta},
k_{\alpha}k_{\beta}k_{\mu},k_{\alpha}k_{\beta}k_{\mu}k_{\nu}\big]\,,
\label{eq:d39}
\end{split}
\end{equation}
which are calculated in App.~1. By means of the $I_{m}^{\sss (\pm)}(p)_{\dots}$-integrals, the $D_{\dots|\dots}^{\sss (\pm),m}$-functions
in momentum space are
\begin{gather}
\hat{D}_{\cdot|\cdot}^{\sss(\pm),m}(p)=\frac{-1}{(2\pi)^{4}}\,\Big[I_{m}^{\sss (\pm)}(p)\Big]\,, \nonumber\\
\hat{D}_{\alpha |\beta}^{\sss (\pm),m}(p)=\frac{+1}{(2\pi)^4}\Big[+p_{\alpha}\,I_{m}^{\sss(\pm)}(p)_{\beta}-
I_{m}^{\sss(\pm)}(p)_{\alpha\beta}\Big]\,,\nonumber\\
\begin{split}
\hat{D}_{\alpha\beta |\mu\nu}^{\sss(\pm),m}(p)=\frac{-1}{(2\pi)^4}\Big[&+p_{\alpha}p_{\beta}\,I_{m}^{\sss(\pm)}(p)_{\mu\nu}
-p_{\alpha}\,I_{m}^{\sss(\pm)}(p)_{\beta\mu\nu}\\
&-p_{\beta}\,I_{m}^{\sss(\pm)}(p)_{\alpha\mu\nu}+I_{m}^{\sss(\pm)}(p)_{\alpha\beta\mu\nu}\Big]\,.
\label{eq:d40}
\end{split}
\end{gather}
Inserting~(\ref{eq:d40}) with Eqs.~(\ref{eq:da6}),~(\ref{eq:da7}),~(\ref{eq:da11}),~(\ref{eq:da15}) and~(\ref{eq:da19}) 
into~(\ref{eq:d37}), then the $a'_{2}$-distribution in momentum space reads
\begin{equation}
\begin{split}
\hat{a}'_{2}(p)^{g\sss SE}_{\alpha\beta\mu\nu}=\frac{-\kappa ^2 \pi}{960 (2\pi)^4}\,\Big[
&+A\:  p_{\alpha}p_{\beta}p_{\mu}p_{\nu} 
 +B \:  p^2\big( p_{\alpha}p_{\beta}\eta_{\mu\nu}+p_{\mu}p_{\nu}\eta_{\alpha\beta}\big)+\\
&+C\:  p^2 \big(p_{\alpha}p_{\mu}\eta_{\beta\nu}+p_{\alpha}p_{\nu}\eta_{\beta\mu}+p_{\beta}p_{\mu}\eta_{\alpha\nu}+
                   p_{\beta}p_{\nu}\eta_{\alpha\mu}\big)+ \\
&+E\:  p^4 \big(\eta_{\alpha\mu}\eta_{\beta\nu}+\eta_{\alpha\nu}\eta_{\beta\mu}\big)\\
&+F\:  p^4 \eta_{\alpha\beta}\eta_{\mu\nu}\Big]\,\hat{d}(p)_{\sss m}^{\sss (+)}\,,
\label{eq:d41}
\end{split}
\end{equation}
with the coefficients
\begin{gather}
A:=-8 -16\,\frac{m^2}{p^2} -48\,\frac{m^4}{p^4}\,,\qquad B:=-4 -8\,\frac{m^2}{p^2} -24\,\frac{m^4}{p^4}\,,\nonumber\\
C:=+1  -8\,\frac{m^2}{p^2} +16\,\frac{m^4}{p^4}\,,\qquad E:=-1 +8\,\frac{m^2}{p^2} -16\,\frac{m^4}{p^4}\,,\nonumber\\
F:=-1 -12\,\frac{m^2}{p^2} +4\,\frac{m^4}{p^4}\,,
\label{eq:d42}
\end{gather}
and the distribution $\hat{d}(p)_{\sss m}^{\sss (\pm)}:=\sqrt{1-\frac{4m^2}{p^2}}\,\Theta(p^2-4m^2)\,\Theta(\pm p^0)$. Performing the 
same calculations for $R'_{2}(x,y)=-T_{1}^{\sss M}(y)T_{1}^{\sss M}(x)$, we obtain
\begin{gather}
R'_{2}(x,y)^{g\sss SE}=:\!h^{\alpha\beta}(x)h^{\mu\nu}(y)\!:\,r'_{2}(x-y)_{\alpha\beta\mu\nu}^{g\sss SE}\,,\nonumber\\
\hat{r}'_{2}(p)_{\alpha\beta\mu\nu}^{g\sss SE}=\frac{-\kappa^2\pi}{960(2\pi)^4}\,
\big[\text{the same as in Eq.~(\ref{eq:d41})}\big]\,\hat{d}(p)_{\sss m}^{\sss (-)}\,.
\label{eq:d44}
\end{gather}
Therefore, with~(\ref{eq:d4}) the causal $D_{2}(x,y)$-distribution reads
\begin{gather}
D _{2}(x,y)^{g\sss SE}=:\!h^{\alpha\beta}(x)h^{\mu\nu}(y)\!:\,d _{2}(x-y)_{\alpha\beta\mu\nu}^{g\sss SE}\,,\nonumber\\
\hat{d}_{2}(p)_{\alpha\beta\mu\nu}^{g\sss SE}=\frac{\kappa^2\pi}{960(2\pi)^4}\,
\big[\text{the same as in Eq.~(\ref{eq:d41})}\big]\,\hat{d}(p)_{\sss m}\,.
\label{eq:d45}
\end{gather}
Here, $\hat{d}(p)_{\sss m}=\hat{d}(p)_{\sss m}^{\sss (+)}-\hat{d}(p)_{\sss m}^{\sss (-)}
=\sqrt{1-\frac{4m^2}{p^2}}\,\Theta(p^2-4m^2)\,\mathrm{sgn}(p^0)$. The $d_{2}$-distribution can be recast into the form
\begin{equation}
\begin{split}
\hat{d}_{2}(p)_{\alpha\beta\mu\nu}^{g\sss SE}&=\sum_{i=1}^{3}\,\hat{d}(p)_{\alpha\beta\mu\nu}^{\sss (i)}\\
&=\underbrace{\frac{\kappa^2\pi}{960(2\pi)^4}}_{=:\Upsilon}\,\Big[\hat{P}(p)_{\alpha\beta\mu\nu}+\frac{m^2}{p^2}\,
\hat{Q}(p)_{\alpha\beta\mu\nu}+\frac{m^4}{p^4}\,\hat{R}(p)_{\alpha\beta\mu\nu}\Big]\,\hat{d}(p)_{\sss m}\,,
\label{eq:d46}
\end{split}
\end{equation}
where the polynomials of degree four are given by their coefficients
\begin{gather}
\hat{P}(p)_{\alpha\beta\mu\nu}=\big[-8,-4,+1,-1,-1\big]\,,\nonumber\\
\hat{Q}(p)_{\alpha\beta\mu\nu}=\big[-16,-8,-8,+8,-12\big]\,,\nonumber\\
\hat{R}(p)_{\alpha\beta\mu\nu}=\big[-48,-24,+16,-16,+4\big]\,,
\label{eq:d47}
\end{gather}
according to the structure given in Eq.~(\ref{eq:d41}).

In the case of massless ($m=0$) matter coupling, that is the first order matter interaction is chosen to be
$T_{1}^{\sss M}(x)=i\frac{\kappa}{2}\,:\!h^{\alpha\beta}(x)\f(x)_{,\alpha}\f(x)_{,\beta}\!:$, then the $d_{2}$-distribution reads
\begin{equation}
\hat{d}_{2}(p)_{\alpha\beta\mu\nu}^{m=0}=\Upsilon\,\hat{P}(p)_{\alpha\beta\mu\nu}\,\Theta(p^2)\,\mathrm{sgn}(p_0)\,.
\label{eq:d47.1}
\end{equation}
Hence, the limit $m\to 0$ of~(\ref{eq:d46}) is feasible without problems, see Eq.~(\ref{eq:d77.1}) for the splitting in the $m=0$ case.

The extension to non-minimally coupled massless matter is also considered. From
\begin{equation}
\mathcal{L}_{\sss M}=\frac{1}{2}\,\sqrt{-g}\,g^{\mu\nu}\,\f_{;\mu}\f_{;\nu}+\frac{1}{12}\,\sqrt{-g}\,R\,\f^2 \,,
\label{eq:d47.2}
\end{equation}
we derive the first order matter coupling
\begin{equation}
T_{1}^{\sss M}(x)=i\,\frac{\kappa}{2}\bigg\{ \frac{2}{3}\,:\!h^{\alpha\beta}\f_{,\alpha}\f_{,\beta}\!:-\frac{1}{6}\,
:\!h \f_{,\sigma}\f_{,\sigma}\!:-\frac{1}{3}\,:\!h^{\alpha\beta}\f \f_{,\alpha \beta}\!:\bigg\}\,,
\label{eq:d47.3}
\end{equation}
which yields~(see~\cite{gri4},~\cite{gri5} for the calculations in the $m=0$ case)
\begin{equation}
\hat{d}_{2}(p)_{\alpha\beta\mu\nu}^{\text{non-min.}}=\Upsilon\,\big[-\frac{12}{9},-\frac{6}{9},1,-1,\frac{6}{9}\big]\,\Theta(p^2)\,
\mathrm{sgn}(p_0)\,.
\label{eq:d47.4}
\end{equation}
The corresponding $t_{2}$-distribution will be given in Eq.~(\ref{eq:d77.2}).

\subsection{Distribution Splitting and UV Finiteness for the Matter Loop}\label{sec:32}

We turn now to the splitting of the $D_{2}$-distribution of Eq.~(\ref{eq:d46}). The leading singular order is four because  the 
polynomials are  of degree four in $p$. But, since these polynomials act in configuration space as derivatives, the essential structure
of the distributions is given by the scalar part. Therefore, neglecting the polynomials, the first, the second
and the third term in~(\ref{eq:d46}) has singular order $0$, $-2$ and $-4$, respectively, due to the inverse powers of $p$.
When discussing the normalization  $N_{2}(x,y)$, we will realize that this was the correct choice. 

As anticipated in Sec.~\ref{sec:21}, a retarded part of $d_{2}$ is obtained in momentum space by the integral~\cite{scha}:
\begin{equation}
\hat{r}_{c}(p_0)=\frac{i}{2\pi}\,p_0^{\omega +1}\,\int_{-\infty}^{+\infty}\!\!dk_0 \,
\frac{\hat{d}(k_0)}{(k_0 -i0)^{\omega +1}(p_0 -k_0+i0)}\,,
\label{eq:d48}
\end{equation}
for $p_{\nu}=(p_0,\bol{0})$, $p_0 >0$. This retarded part  is the so-called `central splitting solution',
because the subtraction point is the origin. 

The behaviour of the first term in~(\ref{eq:d46}) is dictated by a scalar distribution of the form
\begin{equation}
\hat{d}(k)^{\sss (1)}=\sqrt{1-\frac{4m^2}{k^2}}\,\Theta(k^2-4m^2)\,\mathrm{sgn}(k_0)=:  f(k^2)\,\mathrm{sgn}(k_0)\,.
\label{eq:d49}
\end{equation}
Therefore, we have to split $\hat{d}(k)$ with $\omega(\hat{d})=0$, because $\hat{d}(k_{0})\sim constant$ for 
$|k_{0}|\to\infty$ and $k_{\nu}=(k_{0},\bol{0})\in V^{\sss +}$. From~(\ref{eq:d48}) we obtain
\begin{equation}
\begin{split}
\hat{r}_{c}(p_0)
&=\frac{i}{2\pi}\,p_0\,\int_{-\infty}^{+\infty}\!\!dk_0 \,
                            \frac{f(k_0^2)\,\mathrm{sgn}(k_0)}{(k_0-i0)(p_0 -k_0+i0)}\\
&=\frac{i}{2\pi}\,p_0\,\left[ \int_{0}^{\infty} \!\!dk_0\,\frac{f(k_0^2)}{k_0 (p_0-k_0+i0)}-
               \int_{-\infty}^{0}\!\!dk_0\,\frac{f(k_0^2)}{k_0 (p_0-k_0+i0)}\right]\\
&=\frac{i}{2\pi}\,p_0\, \int_{0}^{\infty} \!\!dk_0\,\frac{f(k_0^2)}{k_0}\Big( \frac{1}{p_0-k_0+i0} +\frac{1}{p_0+k_0+i0}\Big)\\
&=\frac{i}{2\pi}\,p_0\, \int_{0}^{\infty} \!\!dk_0\,\frac{f(k_0^2)}{k_0^2}\, \frac{2p_0 k_0}{p_0^2-k_0^2+ip_0 0}\,.
\label{eq:d50}
\end{split}
\end{equation}
With the new variable $s:=k_0^2$ we find
\begin{equation}
\hat{r}_{c}(p_0)=\frac{i}{2\pi}\,p_0^{2}\,\int_{0}^{\infty} \!\!ds\,\frac{f(s)}{s(p_0^2-s+ip_0 0)}\,.
\label{eq:d51}
\end{equation}
Inserting the explicit form of $f(s)$ we have
\begin{equation}
\hat{r}_{c}(p_0)=\frac{i}{2\pi}\,p_0^{2}\,\int_{4m^2}^{\infty}\!\!ds\,\sqrt{1-\frac{4m^2}{s}}\,\frac{1}{s(p_0^2-s+ip_0 0)}\,.
\label{eq:d52}
\end{equation}
We decompose the integral into real and imaginary part according to $(x+i0)^{-1}=P(x^{-1})-i\pi\delta(x)$:
\begin{multline}
\hat{r}_{c}(p_0)=\frac{i}{2\pi}\,p_0^{2}\,P\int_{4m^2}^{\infty}\!\!ds\,\sqrt{1-\frac{4m^2}{s}}\,\frac{1}{s(p_0^2-s)}+\\
+\frac{1}{2}\,\sqrt{1-\frac{4m^2}{p_0^2}}\,\Theta(p_0^2-4m^2)\,\mathrm{sgn}(p_0)\,.
\label{eq:d53}
\end{multline}
The $T_{2}(x,y)$-distribution is obtained from the retarded part $R_{2}(x,y)$  by subtracting $R'_{2}(x,y)$. This subtraction
affects only the numerical distributions. Therefore, subtracting
\begin{equation}
\hat{r}'(p_0)=-\sqrt{1-\frac{4m^2}{p_0^2}}\,\Theta(p_0^2-4m^2)\,\Theta(-p_0)\,,
\label{eq:d54}
\end{equation}
from~(\ref{eq:d53}), we obtain the numerical $\hat{t}$-distribution belonging to $T_{2}(x,y)$:
\begin{equation}
\begin{split}
\hat{t}(p_0)&=\hat{r}_{c}(p_0)-\hat{r}'(p_0)\\
&=\frac{i}{2\pi}\,p_0^2\,P\int_{4m^2}^{\infty}\!\!ds\,\sqrt{1-\frac{4m^2}{s}}\,\frac{1}{s(p_0^2-s)}
+\frac{1}{2}\,\sqrt{1-\frac{4m^2}{p_0^2}}\,\Theta(p_0^2-4m^2)\,,
\label{eq:d55}
\end{split}
\end{equation}
which can be written in the form
\begin{equation}
\hat{t}(p_0)=\frac{i}{2\pi}\,p_0^2\,\int_{4m^2}^{\infty}\!\!ds\,\sqrt{1-\frac{4m^2}{s}}\,\frac{1}{s(p_0^2-s+i0)}\,.
\label{eq:d56}
\end{equation}
This result can be generalized for $p\in V^{\sss +}$ by introducing the `inverse momentum' $q:=4m^2 /p^2$ so that
\begin{equation}
\hat{t}(p)=\frac{i}{2\pi}\,\int_{q}^{\infty}\!\!ds\frac{\sqrt{s(s-q)}}{s^2(1-s+i0)}=:\frac{i}{2\pi}\,\hat{\Pi}(p^2)\,.
\label{eq:d57}
\end{equation}
Note that we  write for simplicity the $p$-dependence instead of the $q$-dependence of the basic integral $\hat{\Pi}$ that remains to
be calculated. Therefore, the splitting of the first term in~(\ref{eq:d46}) and the subtraction of 
$\hat{r}'(p)_{\alpha\beta\mu\nu}^{\sss (1)}$ yields
\begin{equation}
\hat{t}(p)_{\alpha\beta\mu\nu}^{\sss (1)}=i\,\Xi\,\hat{P}(p)_{\alpha\beta\mu\nu}\,\hat{\Pi}(p^2)\,.
\label{eq:d58}
\end{equation}
Here, $\Xi:=\Upsilon/(2\pi)=\kappa^2 \pi /(960 (2\pi)^5)$.

Following the same steps as from Eq.~(\ref{eq:d50}) to Eq.~(\ref{eq:d58}), we find for the second term in~(\ref{eq:d46})
\begin{equation}
\hat{t}(p)_{\alpha\beta\mu\nu}^{\sss (2)}=i\,\Xi\,\frac{m^2}{p^2}\,\hat{Q}(p)_{\alpha\beta\mu\nu}\,\hat{\Pi}(p^2)\,,
\label{eq:d59}
\end{equation}
because in  $\hat{d}(p)_{\alpha\beta\mu\nu}^{\sss (2)}=m^{2}\, \Upsilon\, \hat{Q}(p)_{\alpha\beta\mu\nu}\, \hat{d}(p)^{\sss (2)}$,
the scalar part reads 
\begin{equation}
\hat{d}(p)^{\sss (2)}=\frac{1}{p^2}\,\sqrt{1-\frac{4m^2}{p^2}}\,\Theta(p^2-4m^2)\,\mathrm{sgn}(p_0)\, .
\end{equation}
Analogously, for the third term of~(\ref{eq:d46}) we obtain
\begin{equation}
\hat{t}(p)_{\alpha\beta\mu\nu}^{\sss (3)}=i\,\Xi\,\frac{m^4}{p^4}\,\hat{R}(p)_{\alpha\beta\mu\nu}\,\hat{J}(p^2)\,,
\label{eq:d61}
\end{equation}
because in $\hat{d}(p)_{\alpha\beta\mu\nu}^{\sss (3)}=m^{4}\, \Upsilon\, \hat{R}(p)_{\alpha\beta\mu\nu}\, \hat{d}(p)^{\sss (3)}$,
the scalar part reads
\begin{equation}
\hat{d}(p)^{\sss (3)}=\frac{1}{p^4}\,\sqrt{1-\frac{4m^2}{p^2}}\,\Theta(p^2-4m^2)\,\mathrm{sgn}(p_0)\,,
\end{equation}
with the definition
\begin{equation}
\hat{J}(p^2):=\int_{q}^{\infty}\!\!ds\frac{\sqrt{s(s-q)}}{s^3(1-s+i0)}\,.
\label{eq:d62}
\end{equation}
The difference between $\hat{J}(p^2)$ and  $\hat{\Pi}(p^2)$ lies in the powers of $s$ 
in the denominators in Eqs.~(\ref{eq:d57}) and~(\ref{eq:d62}). This is a consequence of 
splitting these distributions according
to the singular orders of the corresponding scalar distributions.

The two integrals $\hat{\Pi}(p^2)$ and $\hat{J}(p^2)$  have the same structure and the last  can be expressed 
by means  of the first. We decompose $\hat{\Pi}(p^2)$ into real and imaginary part
\begin{equation}
\hat{\Pi}(p^2)=P\int_{q}^{\infty}\!\!ds\,\frac{\sqrt{s(s-q)}}{s^2(1-s)} -i\,\pi\,\sqrt{1-q}\,\Theta(1-q)\,,
\label{eq:d63}
\end{equation}
and concentrate our attention to the principal value part. With the 
substitution\footnote{we choose $x(s)=s+\sqrt{s(s-q)}$, so that $x(s)$ goes from $q$ to $\infty$ for $s$ going also from $q$ to 
$\infty$.} $s(s-q)=(s-x)^2$ the real part of $\hat{\Pi}(p^2)$ reads
\begin{equation}
\begin{split}
\hat{\Pi}_{r}(p^2)&=-2\, P\int_{q}^{\infty}\!\! dx\,\frac{ (x-q)^2}{x^2(x^2-2x+q)}\\
&=-2\, P\int_{q}^{\infty}\!\! dx\,\left\{ \frac{ q}{x^2}+\frac{1-q}{x^2-2x+q}\right\}\,,
\label{eq:d64}
\end{split}
\end{equation}
having factorized the integrand. The $\hat{J}(p^2)$-integral yields a real part
\begin{equation}
\begin{split}
\hat{J}_{r}(p^2)&=-2\, P\int_{q}^{\infty}\!\! dx\,\frac{ (x-q)^2(2x-q)}{x^4(x^2-2x+q)}\\
&=-2\, P\int_{q}^{\infty}\!\! dx\,\left\{ \frac{q-1}{x^2}+\frac{2q}{x^3}-\frac{q^2}{x^4} +\frac{1-q}{x^2-2x+q}\right\}\,.
\label{eq:d65}
\end{split}
\end{equation}
A look at Eqs.~(\ref{eq:d64}) and~(\ref{eq:d65}) enables us to isolate in the expression for $\hat{J}_{r}(p^2)$ 
the terms appearing also in~(\ref{eq:d64}). The others can be easily integrated and we obtain
\begin{equation}
\hat{J}_{r}(p^2)=\frac{2}{3q}+\hat{\Pi}_{r}(p^2)=\frac{p^2}{6m^2}+\hat{\Pi}_{r}(p^2)\,,
\label{eq:d67}
\end{equation}
The imaginary parts have always the same form as in Eq.~(\ref{eq:d63}). Gathering the results 
in~(\ref{eq:d58}),~(\ref{eq:d59}) and~(\ref{eq:d61}) with~(\ref{eq:d67})  we can write the  distribution describing the matter
loop graviton self-energy:
\begin{equation}
\begin{split}
\hat{t}_{2}(p)_{\alpha\beta\mu\nu}^{g\sss SE}&=\sum_{i=1}^{3}\,\hat{t}(p)_{\alpha\beta\mu\nu}^{\sss (i)}\\
&=i\,\Xi\,\bigg[\Big\{ \hat{P}(p)_{\alpha\beta\mu\nu}+\frac{m^2}{p^2}\,\hat{Q}(p)_{\alpha\beta\mu\nu}
+\frac{m^4}{p^4}\,\hat{R}(p_{\alpha\beta\mu\nu})\Big\}\,\hat{\Pi}(p^2)+\\
&\qquad\qquad +\frac{m^2}{6p^2}\,\hat{R}(p)_{\alpha\beta\mu\nu}\bigg]\,.
\label{eq:d68}
\end{split}
\end{equation}
Therefore, the $2$-point operator-valued distribution $T_{2}(x,y)$ for the graviton self-energy reads
\begin{equation}
T_{2}(x,y)^{g\sss SE}=:\!h^{\alpha\beta}(x)h^{\mu\nu}(y)\!:\,t_{2}(x-y)_{\alpha\beta\mu\nu}^{g\sss SE}=
:\!h^{\alpha\beta}(x)h^{\mu\nu}(y)\!:\,i\,\Pi(x-y)_{\alpha\beta\mu\nu}\,,
\label{eq:d69}
\end{equation}
where $\Pi(x-y)_{\alpha\beta\mu\nu}$ is the graviton self-energy tensor. Its Fourier representation is given by
$-i\times$(\ref{eq:d68}).

We still must calculate explicitly the integral representation for $\hat{\Pi}(p^2)$,~(\ref{eq:d57}). There are three different
regimes, depending on the value of $q$. For $q=1$, namely $p^2=4m^2$, we obtain by means of the partial 
decomposition~(\ref{eq:d64}):
\begin{equation}
\hat{\Pi}_{a}(p^2=4m^2)=-2\int_{1}^{\infty}\!\! dx\,\frac{1}{x^2}=-2\,.
\label{eq:d70}
\end{equation}
For $q<1$, namely $p^2 > 4m^2$, the integration of the partial decomposition~(\ref{eq:d64}), taking into account also the imaginary part 
from~(\ref{eq:d63}), yields
\begin{equation}
\hat{\Pi}_{b}(p^2)=-2+\sqrt{1-q}\,\log\left|\frac{q-1-\sqrt{1-q}}{q-1+\sqrt{1-q}}\right| -i\,\pi\,\sqrt{1-q}\,\Theta(1-q)\,.
\label{eq:d71}
\end{equation}
For $q>1$, namely $0<p^2<4m^2$, the integration of~(\ref{eq:d63}) gives
\begin{equation}
\begin{split}
\hat{\Pi}_{c}(p^2)&=-2+2\,\sqrt{q-1}\,\left(\frac{\pi}{2}-\arctan\frac{q-1}{\sqrt{q-1}}\right)\\
&=-2+2\,\sqrt{q-1}\,\arctan\left(\frac{1}{\sqrt{q-1}}\right)\,.
\label{eq:d72}
\end{split}
\end{equation}
Note that these three results are connected by
\begin{equation}
\lim_{p^2\searrow 4m^2}\hat{\Pi}_{b}(p^2)=\hat{\Pi}_{a}(p^2=4m^2)\,,\quad 
\lim_{p^2\nearrow 4m^2}\hat{\Pi}_{c}(p^2)=\hat{\Pi}_{a}(p^2=4m^2)\,.
\label{eq:d73}
\end{equation}
Writing the $p$-dependence explicitly, the final form for $\hat{\Pi}(p^2)$ is
\begin{equation}
\begin{split}
\hat{\Pi}(p^2)=-2 -\Bigg\{&+\sqrt{1-\frac{4m^2}{p^2}}\,\bigg[
\log\left| \frac{1-\sqrt{1-\frac{4m^2}{p^2}}}{1+\sqrt{1-\frac{4m^2}{p^2}}}\right| +i\,\pi\bigg]\,\Theta(p^2-4m^2)+\\
&-2\,\sqrt{\frac{4m^2}{p^2}-1}\,\arctan\frac{1}{\sqrt{\frac{4m^2}{p^2}-1}}\,\Theta(4m^2-p^2)\Bigg\}\,.
\label{eq:d74}
\end{split}
\end{equation}

Two limits of this result will be used in the discussion of the normalization $N_{2}$ of the  $T_{2}$-distribution, 
Sec.~\ref{sec:34}: the limit of $\hat{\Pi}(p^2)$ for $p^2 \searrow 0$ and the limit of $\hat{\Pi}(p^2)/p^2$ for 
$p^2 \searrow 0$, too. In the first case we have
\begin{equation}
\begin{split}
\lim_{p^2\searrow 0}&\hat{\Pi}(p^2)=\lim_{q\to\infty}\hat{\Pi}_{c}(p^2)\\
&=\lim_{q\to\infty}\Big\{-2 +2\sqrt{q-1}\,\Big(\frac{\pi}{2}-\big(\frac{\pi}{2}-\frac{1}{\sqrt{q-1}}+\frac{1}{3\sqrt{q-1}^3}+
          O(\frac{1}{q^{5/2}})\big)\Big)\Big\}\\
&=\lim_{q\to\infty}\Big\{-2 +2-\frac{2}{3(q-1)}+O(\frac{1}{q^{2}})\Big\}=0\,.
\label{eq:d75}
\end{split}
\end{equation}
For the second limit, we obtain
\begin{equation}
\begin{split}
\lim_{p^2\searrow 0}&\frac{\hat{\Pi}(p^2)}{p^2}=\lim_{q\to\infty}\frac{q\,\hat{\Pi}_{c}(p^2)}{4m^2}\\
&=\lim_{q\to\infty}\frac{q}{4m^2}\,\Big\{-2 +2\sqrt{q-1}\,\Big(\frac{1}{\sqrt{q-1}}-\frac{1}{3\sqrt{q-1}^3}+
       O(\frac{1}{q^{5/2}})\Big)\Big\}=\frac{-1}{6m^2}\,.
\raisetag{19mm}\label{eq:d76}
\end{split}
\end{equation}

The last consideration concerns the retarded part in Eq.~(\ref{eq:d53}), given also by~(\ref{eq:d74}) up to the signum-function
in $p_0$: this retarded part is the boundary value of the analytic function
\begin{equation}
\hat{r}(p)^{an}=-2-\sqrt{1-\frac{4m^2}{p^2}}\,\log\frac{\sqrt{1-\frac{4m^2}{p^2}}-1}{\sqrt{1-\frac{4m^2}{p^2}}+1}\,.
\label{eq:d77}
\end{equation}

Summing up the whole calculation, we have found the $2$-point distribution~(\ref{eq:d69}) for the graviton self-energy contribution.
The corresponding tensor, the structure given by~(\ref{eq:d68}) and the integral in~(\ref{eq:d74}), is UV finite and cutoff-free.
 
During the calculation, we never resorted to an {\emph{ad-hoc}} regularization of the expressions 
(for example dimensional regularization as in~\cite{cap3}). This was made possible by using the correct 
starting point, namely Eq.~(\ref{eq:d48}), which is, so to say, a careful multiplication by a step-function 
in the time argument. If this had been done naively, then it would have corresponded to the choice
$\omega =-1$ in~(\ref{eq:d48}), when splitting the first term of Eq.~(\ref{eq:d46}), a choice which is manifestly wrong, being $\omega=0$.

Choosing $\omega=-1$ in Eq.~(\ref{eq:d50}), one obtains a UV logarithmic divergence.

For the sake of completeness, we briefly report also the results in the case of massless matter coupling, Eq.~(\ref{eq:d47.1}),
and in the case of non-minimally coupled massless matter, Eq.~(\ref{eq:d47.4}).

The splitting of the scalar distribution $\Theta(p^2)\,\mathrm{sgn}(p_0)$ requires some modifications if one tries to use the splitting
formula~(\ref{eq:d48}), see~\cite{ym2},~\cite{gri4}. The retarded part is given by $(i/ 2\pi)\log\big((-p^2-i\,p_0 0)/M^2\big)$, so that
the $m=0$ matter self-energy tensor reads
\begin{equation}
\hat{\Pi}(p)_{\alpha\beta\mu\nu}^{m=0}=\Xi\,\hat{P}(p)_{\alpha\beta\mu\nu}\,\log\left( \frac{-p^2 -i0}{M^2}\right)\,,
\label{eq:d77.1}
\end{equation}
where $M>0$ is a scale invariance breaking normalization constant and not a cutoff.

For the non-minimally coupled case we find analogously
\begin{equation}
\hat{\Pi}(p)_{\alpha\beta\mu\nu}^{\text{non-min.}}=\Xi\,\big[-\frac{12}{9},-\frac{6}{9},1,-1,\frac{6}{9}\big]\,
\log\left( \frac{-p^2 -i0}{M^2}\right)\,.
\label{eq:d77.2}
\end{equation}
This graviton self-energy tensor is traceless: $\eta^{\alpha\beta}\,\hat{\Pi}(p)_{\alpha\beta\mu\nu}^{\text{non-min.}}=0$,
because in this case the graviton is coupled to a traceless matter energy-momentum tensor. The latter corresponds to the so-called
`improved' energy-momentum tensor~\cite{calla}. 

In addition, it is transversal:
$p^{\alpha}\,\hat{\Pi}(p)_{\alpha\beta\mu\nu}^{\text{non-min.}}=0$. This property follows from the gauge identity~(\ref{eq:d33}),
namely $b^{\alpha\beta\rho\sigma}p_{\sigma}\hat{\Pi}(p)_{\alpha\beta\mu\nu}=0$, see Sec.~\ref{sec:33}, and from its vanishing trace.

Also in these two cases, we have found UV finite graviton self-energy contributions without introducing counterterms or UV cutoff.
This is in contrast to the calculations carried out for massless scalar matter fields coupled to QG in the background field 
method with  dimensional regularization~\cite{hooft},~\cite{velt}.

\subsection{Graviton Self-Energy Tensor and Perturbative Gauge Invariance}\label{sec:33}

The gauge properties of $T_{2}(x,y)^{g\sss SE}$ are contained in the identity 
$b^{\alpha\beta\rho\sigma}\d_{\sigma}^{x}\Pi(x-y)_{\alpha\beta\mu\nu}=0$, see Eq.~(\ref{eq:d33}). This identity implies the conditions
\begin{equation}
A-2B=0\,,\quad C+E=0\,,\quad B-2E-2F=0\,,
\label{eq:d78}
\end{equation}
for the coefficients of the self-energy tensor. These conditions are satisfied by our result of Eq.~(\ref{eq:d42}) and therefore
$\Pi(x-y)_{\alpha\beta\mu\nu}$ is gauge invariant. This is certainly the case at the level of the $D_{2}(x,y)^{g\sss SE}$-distribution, 
before distribution splitting. In the causal scheme, UV finiteness and gauge invariance are established separately. The latter is not 
used to reach the former. The identity~(\ref{eq:d33}) implies the Slavnov--Ward identities (SWI) for the $2$-point connected Green
function. The latter is defined as
\begin{equation}
\hat{G}(p)_{\alpha\beta\mu\nu}^{\sss [2]}:=b_{\alpha\beta\gamma\delta}\,\hat{D}_{0}^{\sss F}(p)\,
\hat{\Pi}(p)^{\gamma\delta\rho\sigma}\,b_{\rho\sigma\mu\nu}\,\hat{D}_{0}^{\sss F}(p)\,,
\label{eq:d79}
\end{equation}
where  $\hat{D}_{0}^{\sss F}(p)=(2\pi)^{-2}(-p^2-i0)^{-1}$ is the scalar Feynman propagator. The two attached lines represent
free graviton Feynman propagators. The  SWI reads~\cite{cap3},~\cite{capSW}:
\begin{equation}
p^{\alpha}\hat{G}(p)_{\alpha\beta\mu\nu}^{\sss [2]}=0\,,
\label{eq:d80}
\end{equation}
namely that the $2$-point connected Green function is transversal. In term of the coefficients $A,\ldots,F$ as in Eq.~(\ref{eq:d41}) we 
have
\begin{equation}
A-2B=0\,,\quad C+E=0\,,\quad \frac{A}{4}+C-F=0\,.
\label{eq:d81}
\end{equation}
These are equivalent to the conditions~(\ref{eq:d78}). This conclusion is valid also in the massless case, Eqs.~(\ref{eq:d77.1}) 
and~(\ref{eq:d77.2}). 

In~\cite{zaidi}, the gauge invariance of the massless matter loop graviton self-energy tensor is also investigated, but there 
it is not realized that the correct matter coupling is the one in Eq.~(\ref{eq:d15}), namely with the $b$-tensor, when one uses the 
Goldberg variable expansion. This deficiency does not have consequences, if the graviton is coupled to a traceless matter 
energy-momentum tensor as in the non-minimal coupling case, Eq.~(\ref{eq:d47.3}).

The condition of perturbative gauge invariance to second order in the loop graph sector
\begin{multline}
d_{Q}T_{2}(x,y)^{g\sss SE}=\d_{\sigma}^{x}\Big(:\!u^{\rho}(x)h^{\mu\nu}(y)\!:\big[b^{\alpha\beta\rho\sigma}\,
      \Pi(x-y)_{\alpha\beta\mu\nu}\big]\Big) +(x\leftrightarrow y)\\
=\d_{\sigma}^{x}\Big(:\!u^{\rho}(x)h^{\mu\nu}(y)\!:\big[t_{uh}^{\sigma}(x-y)_{\rho\mu\nu}
            +\eta^{\rho\sigma}\,t_{uh}(x-y)_{\mu\nu}\big]\Big)+(x\leftrightarrow y)\,,
\label{eq:d82}
\end{multline}
has been explicitly checked by calculating also the distributions $t_{uh}^{\sigma}(x-y)_{\rho\mu\nu}$ and $t_{uh}(x-y)_{\mu\nu}$
with one $Q$-vertex from  Eq.~(\ref{eq:d29}).

\subsection{Reduction of the Freedom in the  Normalization}\label{sec:34}

We turn now to the normalization  of the $T_{2}$-distribution of Eq.~(\ref{eq:d69}). The total singular order is four because  the 
polynomials are  of degree four in $p$. Therefore, we have to add normalization terms up to the singular order four.

The freedom in the normalization due to  the splitting procedure is contained in the local term $N_{2}(x,y)^{g\sss SE}$:
\begin{equation}
N_{2}(x,y)^{g\sss SE}=:\!h^{\alpha\beta}(x)h^{\mu\nu}(y)\!:\,i\,N(\d_{x},\d_{y})_{\alpha\beta\mu\nu}\,\delta^{\sss (4)}(x-y)\,.
\label{eq:d83}
\end{equation}
From~(\ref{eq:d6}), we can write in momentum space this normalization as a sum of polynomials of degree $2i\le4=\omega$
\begin{equation}
\hat{N}(p)_{\alpha\beta\mu\nu}=\sum_{i=0}^{2} \hat{N}(p)^{\sss (2i)}_{\alpha\beta\mu\nu}\,.
\label{eq:d84}
\end{equation}
Normalization terms with odd $\omega$ are absent due to parity and Lorentz covariance. Gauge invariance
$b^{\alpha\beta\rho\sigma}p_{\sigma}\hat{N}(p)^{\sss (2i)}_{\alpha\beta\mu\nu}=0$, $(i=0,1,2)$ and symmetries reduce the freedom in the
normalization in such a way that the polynomials have to be of the form
\begin{gather}
\hat{N}(p)^{\sss (0)}_{\alpha\beta\mu\nu}=0\,,\qquad  \hat{N}(p)^{\sss (2)}_{\alpha\beta\mu\nu}=
\Xi\big[0,0,-a,a,-a\big]\,\frac{1}{p^2}\,,\nonumber\\
\hat{N}(p)^{\sss (4)}_{\alpha\beta\mu\nu}=\Xi\big[4(b+c),2(b+c),-b,b,c\big]\,,
\label{eq:d85}
\end{gather}
in the usual representation given by Eq.~(\ref{eq:d41}). The constants $a,b,c\in \mathbb{R}$ should be fixed by requiring the 
appropriate mass- and coupling constant-normalizations for the corrections of order $\kappa^2$  to the  graviton propagator. 
Letting formally $g\to 1$ in Eq.~(\ref{eq:d1}), we write the  order $\kappa^2$ corrected propagator as
\begin{multline}
-i\,D(x-y)_{\alpha\beta\mu\nu}^{\sss [2]}=-i\,b_{\alpha\beta\mu\nu}\,D_{0}^{\sss F}(x-y) +\int\!\!d^4 x_{1} d^4 x_{2}\,
       \big(-i\,b_{\alpha\beta\rho\sigma}\,D_{0}^{\sss F}(x-x_{1})\big)\\
\times i\,\big(\Pi(x_{1}-x_{2})^{\rho\sigma\gamma\delta}+
      N(x_{1}-x_{2})^{\rho\sigma\gamma\delta}\big)\,\big(-i\,b_{\gamma\delta\mu\nu}\,D_{0}^{\sss F}(x_{2}-y)\big)\,.
\label{eq:d86}
\end{multline}
In momentum space, this becomes
\begin{equation}
\hat{D}(p)_{\alpha\beta\mu\nu}^{\sss [2]}=\frac{-b_{\alpha\beta\mu\nu}}{(2\pi)^2(p^2 +i0)}+\frac{-1}{(p^2 +i0)}\,
   \underbrace{b_{\alpha\beta\rho\sigma}\,\big(\hat{\Pi}+
           \hat{N}\big)(p)^{\rho\sigma\gamma\delta}\,b_{\gamma\delta\mu\nu}}_{=:\tilde{\Pi}(p)_{\alpha\beta\mu\nu}}\,
\frac{-1}{(p^2 +i0)}\,.
\label{eq:d87}
\end{equation}
After a little work, we find in the form of Eq.~(\ref{eq:d41})
\begin{equation}
\tilde{\Pi}(p)_{\alpha\beta\mu\nu}=\Xi\,\big[f_{\sss A}(p^2),f_{\sss B}(p^2) ,f_{\sss C}(p^2) ,f_{\sss E}(p^2) ,f_{\sss F}(p^2) \big]\,,
\label{eq:d88}
\end{equation}
with
\begin{equation}
\begin{split}
f_{\sss A}(p^2)&=\left(-8 -16\,\frac{m^2}{p^2} -48\,\frac{m^4}{p^4}\right)\,\hat{\Pi}(p^2)-\frac{48m^2}{6p^2}+4(b+c)\,,\\
f_{\sss B}(p^2)&=\left(+6 +32\,\frac{m^2}{p^2} +16\,\frac{m^4}{p^4}\right)\,\hat{\Pi}(p^2)+
                           \frac{16m^2}{6p^2}-2b-4c+\frac{2a}{p^2}\\
               &=-f_{\sss F}(p^2)\,,\\
f_{\sss C}(p^2)&=\left(+1 -8\,\frac{m^2}{p^2} +16\,\frac{m^4}{p^4}\right)\,\hat{\Pi}(p^2)+
                           \frac{16m^2}{6p^2}-b -\frac{a}{p^2}=-f_{\sss E}(p^2)\,.
\label{eq:d89}
\end{split}
\end{equation}
With the formula (if $g(p^2)\sim\kappa^2$)
\begin{equation}
\frac{1}{-p^2-i0} \to \frac{1}{-p^2-i0}+\frac{1}{-p^2-i0}\,g(p^2)\,\frac{1}{-p^2-i0}=\frac{1}{-p^2-i0-g(p^2)} +O(\kappa^4)\,,
\label{eq:d90}
\end{equation}
we obtain the order $\kappa^2$ corrected graviton propagator
\begin{equation}
\begin{split}
\hat{D}(p)_{\alpha\beta\mu\nu}^{\sss [2]}=\frac{1}{(2\pi)^2}\,\bigg[&+\frac{1}{2}\,\left(\eta_{\alpha\mu}\eta_{\beta\nu}+
\eta_{\alpha\nu}\eta_{\beta\mu}\right)\,\frac{1}{-p^2-i0-2(2\pi)^2\Xi\,p^4\,f_{\sss E}(p^2)}+\\
&-\frac{1}{2}\left(\eta_{\alpha\beta}\eta_{\mu\nu}\right)\,\frac{1}{-p^2-i0+2(2\pi)^2\Xi\,p^4\,f_{\sss F}(p^2)}\bigg]+\ldots\,,
\label{eq:d91}
\end{split}
\end{equation}
where non-contributing terms between conserved matter energy-momentum tensors have been neglected. The corrected propagator above has the 
correct limit $\lim_{\kappa\to 0}\hat{D}(p)_{\alpha\beta\mu\nu}^{\sss [2]}=b_{\alpha\beta\mu\nu}\hat{D}_{0}^{\sss F}(p)$.

Mass normalization (the graviton mass remains zero under quantum corrections) yields
\begin{equation}
p^4\,f_{\sss E}(p^2)\Big|_{p^2=0}=0=p^4\,f_{\sss F}(p^2)\Big|_{p^2=0}\,.
\label{eq:d92}
\end{equation}
Since $\hat{\Pi}(p^2=0)=0$ from Eq.~(\ref{eq:d75}), these conditions always hold.

Coupling constant normalization ($\kappa$ is not shifted by the quantum corrections) implies
\begin{equation}
p^2\,f_{\sss E}(p^2)\Big|_{p^2=0}=0=p^2\,f_{\sss F}(p^2)\Big|_{p^2=0}\,.
\label{eq:d93}
\end{equation}
Analysis of the first condition
\begin{equation}
p^2\,f_{\sss E}(p^2)\Big|_{p^2=0}=-16m^4\,\underbrace{\frac{\hat{\Pi}(p^2)}{p^2}\bigg|_{p^2=0}}_{=-1/6m^2} -\frac{16m^2}{6}+a=0   \,,
\label{eq:d94}
\end{equation}
yields $a=0$. Analysis of the second condition
\begin{equation}
p^2\,f_{\sss F}(p^2)\Big|_{p^2=0}=-16m^4\,\underbrace{\frac{\hat{\Pi}(p^2)}{p^2}\bigg|_{p^2=0}}_{=-1/6m^2} -\frac{16m^2}{6}-2a=0   \,,
\label{eq:d95}
\end{equation}
yields also  $a=0$. Decisive is the compensation between the first two terms in~(\ref{eq:d94}) and~(\ref{eq:d95}). This is due to the
presence of the term $\frac{m^2}{6p^2}\hat{R}(p)_{\alpha\beta\mu\nu}$ in Eq.~(\ref{eq:d68}).

A remark about the splitting: if we had split the distributions $\hat{d}_{2}(p)_{\alpha\beta\mu\nu}^{\sss (i)}$, $i=2,3$ in 
Eq.~(\ref{eq:d46}) according to their `true' singular orders, namely $2$ and $0$ (because of the presence of the polynomials), 
respectively, then the term 
$\frac{m^2}{6p^2}\hat{R}(p)_{\alpha\beta\mu\nu}$ would have been missing from~(\ref{eq:d68}). Working out the consequences for what 
concerns the normalization question, Eq.~(\ref{eq:d94}) would have required the choice $a=-8m^2/3$, whereas Eq.~(\ref{eq:d95}) the choice
$a=4m^2/3$. This would have meant the impossibility of a consistent normalization. Therefore, the splitting, as carried out in 
Sec.~\ref{sec:32}, is justified.

The origin of the above mentioned problem lies in the fact that the central splitting solution~(\ref{eq:d48}) is not applicable in
that case and one has to choose a subtraction point different from the origin.

The remaining constants $b$ and $c$ are not fixed by these requirements. The total graviton self-energy tensor including its  
normalization has then the form
\begin{equation}
\begin{split}
\hat{\Pi}(p)^{\text{tot}}_{\alpha\beta\mu\nu}&=\Xi\,\bigg[\Big\{ \hat{P}(p)_{\alpha\beta\mu\nu}+\frac{m^2}{p^2}\,
\hat{Q}(p)_{\alpha\beta\mu\nu}+\frac{m^4}{p^4}\,\hat{R}(p_{\alpha\beta\mu\nu})\Big\}\,\hat{\Pi}(p^2)+\\
&\qquad\qquad+\frac{m^2}{6p^2}\,\hat{R}(p)_{\alpha\beta\mu\nu}+\sum_{i=1}^{2}\,z_{i}\,\hat{Z}(p)_{\alpha\beta\mu\nu}^{\sss (i)}\bigg]\,,
\label{eq:d96}
\end{split}
\end{equation}
where 
$z_{i}\in \mathbb{R}$, $i=1,2$ are still undetermined constants. The $\hat{Z}(p)_{\alpha\beta\mu\nu}^{\sss (i)}$'s are basis elements 
in the
two-dimensional space of gauge invariant polynomials of degree four. They can be chosen to be:
$\hat{Z}(p)_{\alpha\beta\mu\nu}^{\sss (1)}=[4,2,-1,1,0]$ and $\hat{Z}(p)_{\alpha\beta\mu\nu}^{\sss (2)}=[4,2,0,0,1]$.

Analysis of the issue of  normalization  with the method used in~\cite{gri4},~\cite{gri5} leads to the same conclusions.

\section{Matter Self-Energy}\label{sec:4}
\setcounter{equation}{0}

The $2$-point distribution describing the matter self-energy graph is not interesting from the point of view of its gauge properties.
However, the calculation of the corresponding distribution is carried out to show its UV finiteness.

\subsection{Causal \boldmath{$D_{2}(x,y)$}-Distribution and Distribution Splitting}\label{sec:41}

The $D_{2}$-distribution is here obtained by performing one matter field contraction and one graviton contraction~(\ref{eq:d34}).
The result reads:
\begin{equation}
\begin{split}
D_{2}(x,y)^{m\sss SE}=&+:\!\f(x)_{,\alpha}\f(y)_{,\alpha}\!:\,
      \Big\{ \Big[\frac{\kappa^2 m^2}{2}\big(C_{\cdot|\cdot}^{\sss (+)} -C_{\cdot|\cdot}^{\sss (-)}\big)(x-y)\Big]\ =:d_{a}(x-y)\Big\}\\
&+:\!\f(x)\f(y)_{,\alpha}\!:\,
\Big\{\Big[\frac{-\kappa^2 m^2}{2}\big(C_{\cdot|\alpha}^{\sss(+)}-C_{\cdot|\alpha}^{\sss(-)}\big)(x-y)\Big]\ =:d_{b}(x-y)^{\alpha}\Big\}\\
&+:\!\f(x)_{,\alpha}\f(y)\!:\,
 \Big\{\Big[\frac{\kappa^2 m^2}{2}\big(C_{\cdot|\alpha}^{\sss(+)}-C_{\cdot|\alpha}^{\sss(-)}\big)(x-y)\Big]\ =:d_{c}(x-y)^{\alpha}\Big\}\\
&+:\!\f(x)\f(y)\!:\,
  \Big\{\Big[-\kappa^2 m^4 \big(C_{\cdot|\cdot}^{\sss (+)} -C_{\cdot|\cdot}^{\sss (-)}\big)(x-y)\Big]=:d_{d}(x-y)\Big\}
\label{eq:d97}
\end{split}
\end{equation}
where the $C^{\sss (\pm)}_{\ldots}$-functions are defined by
\begin{gather}
C^{\sss (\pm)}_{\cdot|\cdot}(x):=D_{0}^{\sss (\pm)}(x)\cdot D_{m}^{\sss (\pm)}(x)\,,\nonumber\\
C^{\sss (\pm)}_{\cdot|\alpha}(x):=D_{0}^{\sss (\pm)}(x)\cdot\d_{\alpha}^{x} D_{m}^{\sss (\pm)}(x)\,.
\label{eq:d98}
\end{gather}
These products are calculated in momentum space, see App.~2, so that the distributions $d_{i}$, $i=a,b,c,d$ read
\begin{gather}
\hat{d}_{a}(p)=\frac{-\kappa^2 m^2 \pi}{4(2\pi)^4}\,\left(1-\frac{m^2}{p^2}\right)\,\Theta(p^2-m^2)\,\mathrm{sgn}(p_0)\,,\nonumber\\
\hat{d}_{b}(p)^{\alpha}=\frac{-i\kappa^2 m^2 \pi}{8(2\pi)^4}\,\left(1-\frac{m^4}{p^4}\right)\,p^{\alpha}\,
                    \Theta(p^2-m^2)\,\mathrm{sgn}(p_0)=-\hat{d}_{c}(p)^{\alpha}\,,\nonumber\\
\hat{d}_{d}(p)=\frac{\kappa^2 m^4 \pi}{2(2\pi)^4}\,\left(1-\frac{m^2}{p^2}\right)\,\Theta(p^2-m^2)\,\mathrm{sgn}(p_0)\,.
\label{eq:d99}
\end{gather}
From power-counting arguments, one could expect that $\hat{d}_{a}$ behaves as $p^2$ for large momenta. This is not the case, because
the wave equation $(\Box + m^2)D_{m}^{\sss (\pm)}(x)=0$ lowers the power of $p$ coming from the product of contractions.
In order to shorten the calculation, we bring $D_{2}(x,y)^{m\sss SE}$ into the form
\begin{equation}
D_{2}(x,y)^{m\sss SE}=:\!\f(x)\f(y)\!:\,d_{2}(x-y)^{m\sss SE}+\text{divergences}\,,
\label{eq:d100}
\end{equation}
with
\begin{equation}
\begin{split}
\hat{d}_{2}(p)^{m\sss SE}=&+p^2\,\hat{d}_{a}(p) -i\,p_{\alpha}\,\hat{d}_{b}(p)^{\alpha} +i\,p_{\alpha}\,\hat{d}_{c}(p)^{\alpha}+
  \hat{d}_{d}(p)\\
&=\underbrace{\frac{-\kappa^2 m^2 \pi}{2(2\pi)^4}}_{=:\Gamma}\,\left[ p^2-\frac{3m^2}{2}+\frac{m^4}{2p^2}\right]\,
   \Theta(p^2-m^2)\,\mathrm{sgn}(p_0)\,.
\label{eq:d101}
\end{split}
\end{equation}
Truly, this simplification can only be made for the corresponding $T_{2}$-distribution, because divergences do not contribute in the 
adiabatic limit $g\to 1$ of Eq.~(\ref{eq:d1}). Therefore, one should split the four distributions $d_{i}$, $i=a,b,c,d$ separately and then
recast the $t_{i}$'s  in a form similar to Eq.~(\ref{eq:d100}) for the $T_{2}$-distribution. The final result would be the same.
Note that in the $m=0$ case, $D_{2}(x,y)^{m\sss SE}=0$.

The splitting of~(\ref{eq:d101}) is accomplished by means of the splitting formula~(\ref{eq:d48}) with $\omega(d_{2}^{m\sss SE})=2$:
\begin{equation}
\hat{r}_{c}(p_0)^{m\sss SE}=\frac{i}{2\pi}\,p_0^3\,\int_{-\infty}^{+\infty}\!\!dk_0\,\frac{\hat{d}_{2}(k_0)^{m\sss SE}}{
(k_0-i0)^3 (p_0 -k_0 +i0)}\,,
\label{eq:d102}
\end{equation}
for $p_{\nu}=(p_0,\bol{0})$, $p_0 >0$. The retarded part then reads
\begin{equation}
\hat{r}_{c}(p_0)^{m\sss SE}=\frac{i\Gamma}{2\pi}\,p_0^3\,\int_{-\infty}^{+\infty}\!\!dk_0\,\frac{\Theta(k_0^2-m^2)\,\mathrm{sgn}(k_0)}
{k_0 (p_0-k_0 +i0)}\,\left\{1-\frac{3m^2}{2k_0^2}+\frac{m^4}{2k_0^4}\right\}\,.
\label{eq:d103}
\end{equation}
With $s=k_0^2$, $ds=2k_0 dk_0$, we obtain
\begin{equation}
\hat{r}_{c}(p_0)^{m\sss SE}=\frac{i\Gamma}{2\pi}\,p_0^4\,\int_{0}^{\infty}\!\!ds\,\frac{\Theta(s-m^2)}
{(p_0^2-s +ip_0 0)}\,\left\{\frac{1}{s}-\frac{3m^2}{2s^2}+\frac{m^4}{2s^3}\right\}\,,
\label{eq:d104}
\end{equation}
which can be decomposed into real and imaginary part:
\begin{multline}
\hat{r}_{c}(p_0)^{m\sss SE}=\frac{i\Gamma}{2\pi}\,p_0^4\,P\int_{0}^{\infty}\!\!ds\,\frac{\Theta(s-m^2)}
{(p_0^2-s)}\,\left\{\frac{1}{s}-\frac{3m^2}{2s^2}+\frac{m^4}{2s^3}\right\}+\\
+\frac{\Gamma}{2}\,\Theta(p_0^2 -m^2)\,\mathrm{sgn}(p_0)\,\left\{p_0^2-\frac{3m^2}{2}+\frac{m^4}{2p_0^2}\right\} \,.
\label{eq:d105}
\end{multline}
Subtracting the distribution
\begin{equation}
\hat{r}'_{2}(p_0)^{m\sss SE}=-\Gamma\,\left[ p_0^2-\frac{3m^2}{2}+\frac{m^4}{2p_0^2}\right]\,\Theta(p_0^2 -m^2)\,\Theta(-p_0)\,,
\label{eq:d106}
\end{equation}
(coming from $R'_{2}(x,y)^{m\sss SE}$) from $\hat{r}_{c}(p_0)^{m\sss SE}$, we find the $2$-point distribution
\begin{equation}
\begin{split}
\hat{t}_{2}(p_0)^{m\sss SE}&=\hat{r}_{c}(p_0)^{m\sss SE}-\hat{r}'_{2}(p_0)^{m\sss SE}\\
  &=\frac{i\Gamma}{2\pi}\,p_0^4\,\int_{0}^{\infty}\!\!ds\,\frac{\Theta(s-m^2)}
{(p_0^2-s +i 0)}\,\left\{\frac{1}{s}-\frac{3m^2}{2s^2}+\frac{m^4}{2s^3}\right\}=:i\,\hat{\Sigma}(p_0)\,.
\label{eq:d107}
\end{split}
\end{equation}

As a next task, we compute the integral of the principal value part of~(\ref{eq:d107}), denoted by $X(p_0)$:
\begin{equation}
\begin{split}
X(p_0)&=\frac{-i\Gamma}{4\pi}\,p_0^4\,P\int_{m^2}^{\infty}\!\!ds\,\frac{2s^2 -3m^2 s +m^4}{s^3(s-p_0^2)}\\
&=\frac{-i\Gamma}{4\pi}\,p_0^4\,P\int_{m^2}^{\infty}\!\!ds\,\left\{\frac{\alpha}{s}+\frac{\beta}{s^2}+\frac{\gamma}{s^3}+
                  \frac{-\alpha}{s-p_0^2}\right\}\,,
\label{eq:d108}
\end{split}
\end{equation}
with 
\begin{equation}
\alpha :=\frac{3m^2}{p_0^4}-\frac{m^4}{p_0^6}-\frac{2}{p_0^2}\,,\quad \beta:=\frac{3m^2}{p_0^2} -\frac{m^4}{p_0^4}\,,\quad
\gamma :=-\frac{m^4}{p_0^2}\,.
\label{eq:d109}
\end{equation}
Integration of the partial fractions in~(\ref{eq:d108}) yields
\begin{equation}
X(p_0)=\frac{i\Gamma}{2\pi}\,\left[\left(p_0^2-\frac{3m^2}{2} +\frac{m^4}{2p_0^2}\right)\,\log\left| \frac{p_0^2 -m^2}{m^2}\right|
+\frac{m^2}{2}-\frac{5p_0^2}{4}\right]\,.
\label{eq:d110}
\end{equation}

\subsection{Matter Self-Energy Two-Point Distribution and Freedom in its Normalization}\label{sec:42}

From~(\ref{eq:d100}) with~(\ref{eq:d107}) and~(\ref{eq:d110}), we can derive the $2$-point distribution 
for the matter self-energy
\begin{equation}
T_{2}(x,y)^{m\sss SE}=:\!\f(x)\f(y)\!:\,i\,\Sigma(x-y)\,,
\label{eq:d111}
\end{equation}
and in an arbitrary Lorentz system, the matter self-energy distribution  for $p\in V^{\sss +}$ is
\begin{equation}
\hat{\Sigma}(p)=\frac{\Gamma}{2\pi}\left[\left(p^2-\frac{3m^2}{2} +\frac{m^4}{2p^2}\right)
\left[\log\left| \frac{p^2 -m^2}{m^2}\right|-i\pi\Theta(p^2 -m^2)\right]+\frac{m^2}{2}-\frac{5p^2}{4}\right]
\label{eq:d112}
\end{equation}
The obtained loop contribution is UV finite and cutoff-free due to the causal scheme. 

This loop contribution was also calculated in~\cite{shiekh} within the operator regularization scheme. 
Parts of their result agree
with our expression in~(\ref{eq:d112}), whereas differences concern the explicit presence of parameters, 
of other $p$-dependent
logarithms  and terms which go as $p^4$ in their expression for $\hat{\Sigma}(p)$. In the causal scheme, 
these latter cannot appear, 
because the singular order remains the same after distribution splitting.

The retarded part in~(\ref{eq:d105}) is the boundary value of the analytic function of complex momentum $p+i\eta$, 
$\eta=(\epsilon,\bol{0})$, $\epsilon >0$:
\begin{equation}
\hat{r}(p)^{an}=\frac{i\Gamma}{2\pi}\,\left[\left(p^2-\frac{3m^2}{2} +\frac{m^4}{2p^2}\right)\,\log\left(1-\frac{m^2}{p^2}\right)
+\frac{m^2}{2}-\frac{5p^2}{4}\right]\,.
\label{eq:d113}
\end{equation}
Having split $d_{2}^{m\sss SE}$ with $\omega=2$, an ambiguity in the  normalization of  $\hat{\Sigma}(p)$ of the type
\begin{equation}
\hat{N}(p)=\frac{\Gamma}{2\pi}\,\big(c_{0}+c_{2}\,p^2\big)\,,
\label{eq:d114}
\end{equation}
must be taken into account. In order to fix the constants $c_{0}$ and $c_{2}$, radiative corrections to the matter Feynman propagator
by matter self-energy loops are considered: letting formally $g\to 1$, they are of the form
\begin{multline}
-i\,D_{m}^{\sss F}(x-y)+\int\!\!d^4 x_{1} d^4 x_{2}\,\big(-i\,D_{m}^{\sss F}(x-x_{1})\big)\\
\times i\,\big[\Sigma(x_{1}-x_{2}) +   N(x_{1}-x_{2})\big]\,\big(-i\,D_{m}^{\sss F}(x_{2}-y)\big)+\ldots\,.
\label{eq:d115}
\end{multline}
In momentum space the series becomes 
\begin{multline}
\hat{D}_{m}^{\sss F}(p) +\hat{D}_{m}^{\sss F}(p)\,(2\pi)^4\, \big( \hat{\Sigma}(p) +\hat{N}(p)\big)\,\hat{D}_{m}^{\sss F}(p)
+\hat{D}_{m}^{\sss F}(p)\times \\ \times(2\pi)^4\,\big( \hat{\Sigma}(p) +\hat{N}(p)\big)\,\hat{D}_{m}^{\sss F}(p)\,(2\pi)^4 \,
\big( \hat{\Sigma}(p) +\hat{N}(p)\big)\,\hat{D}_{m}^{\sss F}(p)+\ldots=:\hat{\Sigma}(p)^{\text{tot}}\,,
\raisetag{12mm}\label{eq:d116}
\end{multline}
where $\hat{D}_{m}^{\sss F}(p)=(2\pi)^{-2}(-p^2 +m^2 -i0)^{-1}$.  The geometric series in~(\ref{eq:d116}) leads to
\begin{equation}
\begin{split}
\hat{\Sigma}(p)^{\text{tot}}&=\hat{D}_{m}^{\sss F}(p)\,\Big(1+(2\pi)^4\,\big(\hat{\Sigma}(p)+\hat{N}(p)\big)\,
              \hat{\Sigma}(p)^{\text{tot}}\Big)\\
&=\frac{1}{(2\pi)^2}\,\frac{1}{-p^2 +m^2 -i0 -(2\pi)^2\,\big(\hat{\Sigma}(p)+\hat{N}(p)\big)}\,.
\label{eq:d117}
\end{split}
\end{equation}
Mass corrections are avoided by requiring
\begin{equation}
\big[\hat{\Sigma}(p)+\hat{N}(p)\big]\Big|_{p^2=m^2} =0\,.
\label{eq:d118}
\end{equation}
This implies
\begin{equation}
c_{0}=m^2\left(\frac{3}{4}-c_{2}\right)\,,
\label{eq:d119}
\end{equation}
so that the matter self-energy distribution including its normalization reads
\begin{multline}
\hat{\Sigma}(p)+\hat{N}(p)=\frac{\Gamma}{2\pi}\,\bigg[\big(p^2-\frac{3m^2}{2} +\frac{m^4}{2p^2}\big)\,
\Big[\log\left| \frac{p^2 -m^2}{m^2}\right|-i\,\pi\,\Theta(p^2 -m^2)\Big]+\\+(p^2-m^2)\,(c_{2}-\frac{5}{4})\bigg]\,.
\label{eq:d120}
\end{multline}
For the remaining constant $c_{2}$, we cannot provide here a value. A restriction should come from the vertex corrections 
$\hat{\Lambda}(p,q)_{\alpha\beta}$ to third order in the $3$-point distribution $T_{3}(x,y,z)=:\f(x)\f(y) h^{\alpha\beta}(z)\!:
\Lambda(x,y,z)_{\alpha\beta}$.

\section{Perturbative Gauge Invariance to Second Order in the Tree Graph Sector}\label{sec:5}
\setcounter{equation}{0}

In this section we show that the condition of perturbative gauge invariance to second order in the tree 
graph sector generates a local quartic interaction of the form $\sim\kappa^2 :\!hh\f\f\!:\delta^{\sss (4)}(x-y)$, 
which agrees with the second order $\mathcal{L}_{\sss M}^{\sss (2)}$ in the expansion of the matter Lagrangian 
density $\mathcal{L}_{\sss M}$, Eq.~(\ref{eq:d26}).

\subsection{Methodology}\label{sec:51}

Since $R'_{2}(x,y)$ is trivially gauge invariant due to its definition and to the  gauge invariance of $T_{1}(x)$,
instead of Eq.~(\ref{eq:d22}) for $n=2$, we have to examine whether
\begin{multline}
d_{Q}R_{2}(x,y)+d_{Q}N_{2}(x,y)=\d_{\nu}^{x}R_{2/1}^{\nu}(x,y)+\d_{\nu}^{y}R_{2/2}^{\nu}(x,y)+\\
+ \d_{\nu}^{x}N_{2/1}^{\nu}(x,y)+\d_{\nu}^{y}N_{2/2}^{\nu}(x,y)
\label{eq:d121}
\end{multline}
can be satisfied by a suitable choice of the free constants in the normalization terms $N_{2}$, $N_{2/1}^{\nu}$ and $N_{2/2}^{\nu}$
of the retarded parts $R_{2}$, $R_{2/1}^{\nu}$ and $R_{2/2}^{\nu}$. Here, $R_{2/1}^{\nu}$ and $R_{2/2}^{\nu}$ are the retarded 
distributions obtained by splitting the inductively constructed distributions
\begin{equation}
D_{2/1}^{\nu}(x,y):=\big[T^{\nu}_{1/1}(x),T_{1}(y)\big]\quad \text{and}\quad D_{2/2}^{\nu}(x,y):=\big[T_{1}(x),T^{\nu}_{1/1}(y)\big]\,.
\label{eq:d122}
\end{equation}
This procedure has been described in~\cite{scho1} for pure QG. It turned out that Eq.~(\ref{eq:d121}) can be spoiled by terms
with point support  $\sim\delta^{\sss (4)}(x-y)$, only. $N_{2}$, $N_{2/1}^{\nu}$ and $N_{2/2}^{\nu}$ are
by definition local terms, but there is another source of `local anomalies', namely the splitting procedure for tree graphs: in the
inductive construction of $R_{2/1}^{\nu}$, the $Q$-vertex $T_{1/1}^{\nu}$ can give rise to expressions of the form
\begin{equation}
D_{2/1}^{\nu}(x,y)=:\!\mathcal{O}(x,y)\!:\,\d^{\nu}_{x} D_{m}(x-y)\,.
\label{eq:d123}
\end{equation}
$:\!\mathcal{O}(x,y)\!:$ is a normally ordered product of four fields, being the other two fields in the 
commutator~(\ref{eq:d122}). $m$ can also be zero, if the commutator contains two graviton fields. The retarded part is simply
\begin{equation}
R_{2/1}^{\nu}(x,y)=:\!\mathcal{O}(x,y)\!:\,\d^{\nu}_{x} D^{\text{ret}}_{m}(x-y)\,,
\label{eq:d124}
\end{equation}
because the trivial splitting for tree graphs follows from $D_{m}=D_{m}^{\text{ret}}-D_{m}^{\text{av}}$ with the correct support 
properties as explained in Sec.~\ref{sec:21}. Applying the derivative $\d_{\nu}^{x}$, which forms the divergence in 
$\d_{\nu}^{x}R_{2/1}^{\nu}$, we get
\begin{equation}
:\!\mathcal{O}(x,y)\!:\,\Box_{x} D^{\text{ret}}_{m}(x-y)=:\!\mathcal{O}(x,y)\!:\,\big[\delta^{\sss (4)}(x-y) 
-m^2\,D_{m}^{\text{ret}}\big]\,.
\label{eq:d125}
\end{equation}
The expression $:\!\mathcal{O}(x,y)\!:\,\delta^{\sss (4)}(x-y)$ is a local anomaly. A corresponding mechanism works 
also for
$R_{2/2}^{\nu}$. 

We define by $\mathrm{an}(\d_{\nu}^{x}R_{2/1}^{\nu}+\d_{\nu}^{y}R_{2/2}^{\nu})$ the set of all local anomalies
generated by the described mechanism. 

Following~\cite{scho1}, gauge invariance is preserved if we can choose
$N_{2}$, $N_{2/1}^{\nu}$ and $N_{2/2}^{\nu}$ so that the condition 
\begin{equation}
d_{Q}N_{2}(x,y)=\mathrm{an}(\d_{\nu}^{x} R_{2/1}^{\nu}+\d_{\nu}^{y} R_{2/2}^{\nu})+\d_{\nu}^{x} N_{2/1}^{\nu}(x,y)
+\d_{\nu}^{y} N_{2/2}^{\nu}(x,y)
\label{eq:d126}
\end{equation}
involving the local terms of~(\ref{eq:d121}) is satisfied. 

Note that in QG coupled to matter  $d_{Q}R_{2}$ does not generate local terms with matter fields involved, namely of the type 
$:\!uh\f\f\!:$. This is in contrast to the much more involved  pure QG case.

At this point we realize that it is not sufficient to consider $T_{1}^{\sss M}$ only. Also the graviton and ghost first order couplings
$T_{1}^{h}$ and $T_{1}^{u}$ have to be taken into account, because they yield also local anomalies with external operators $\sim:\!uh\f\f\!:$  
when splitting the commutators of~(\ref{eq:d122}).

Using a  simplified notation which keeps track of the structure of the coupling only, gauge invariance to first order then becomes
\begin{equation}
d_{Q}\big(\underbrace{:\!hhh\!:+:\!\tilde{u}hu\!:+:\!h\f\f\!:}_{T_{1}^{h+u+\sss M}}\big)=
\d_{\nu}^{x}\big( \underbrace{:\!\{uhh\}^{\nu}\!:+:\!\{\tilde{u}uu\}^{\nu}\!:
+:\!\{u\f\f\}^{\nu}\!:}_{T_{1/1}^{\nu\:h+u}+T_{1/1}^{\nu\:\sss M}}\big)\,.
\label{eq:d127}
\end{equation}
With the `extended' $Q$-vertex $T_{1/1}^{\nu\: h+u} + T_{1/1}^{\nu\: \sss M}$, we decompose the commutator defining $D_{2/1}^{\nu}$
in the following way:
\begin{equation}
\begin{split}
D_{2/1}^{\nu}(x,y)=&\quad \text{pure QG sector}+\big[ :\!\{uhh\}^{\nu}\!:, :\!h\f\f\!:\big] +\\
&+\big[ :\!\{u\f\f\}^{\nu}\!:,:\!\tilde{u}hu\!:\big]+\big[ :\!\{u\f\f\}^{\nu}\!:, :\!h\f\f\!:\big]\,.
\label{eq:d128}
\end{split}
\end{equation}
In the pure QG sector, which involves terms of the type $:\!uhhh\!:$ and $:\!\tilde{u}uuh\!:$, perturbative gauge invariance has  
been shown in~\cite{scho1}.

\subsection{Explicit Calculations}\label{sec:52}

Our task consists in the investigation of the three remaining sectors in Eq.~(\ref{eq:d128}) in which matter and 
graviton fields are mixed together. We denote the three commutators in~(\ref{eq:d128}) as the graviton-, ghost-, 
and matter sector, respectively.

Performing one contraction they lead to
\begin{equation}
\begin{split}
D_{2/1}^{\nu\:h}(x,y)&=\big(:\!\underbrace{uh}_{x}\underbrace{\f\f}_{y}\!:\,\big[h(x),h(y)\big]\big)^{\nu}\,,\\
D_{2/1}^{\nu\:u}(x,y)&=\big(:\!\underbrace{\f\f}_{x}\underbrace{hu}_{y}\!:\,\big[u(x),\tilde{u}(y)\big]\big)^{\nu}\,,\\
D_{2/1}^{\nu\:\sss M}(x,y)&=\big(:\!\underbrace{u\f}_{x}\underbrace{h\f}_{y}\!:\,\big[\f(x),\f(y)\big]\big)^{\nu}\,,
\label{eq:d129}
\end{split}
\end{equation}
respectively. By considering the explicit form of $T_{1}^{u}$ in~\cite{scho1} and $T_{1/1}^{\nu\:\sss M}$ in~(\ref{eq:d21}), 
we find that no local anomalies arise in the ghost sector.

Let us compute the local anomalies in the graviton sector. From the expression for $T_{1/1}^{\nu\: h+u}$ in~\cite{scho1} we isolate
the terms that generate local anomalies according to the mechanism described in Sec.~\ref{sec:51}. They are
\begin{equation}
\begin{split}
T_{1/1}^{\nu\: h+u}=\kappa\,\Big\{
&+\frac{1}{2}:\!u^{\mu}_{,\mu}h^{\rho\sigma}h^{\rho\sigma}_{\:,\nu}\!: 
+\frac{1}{2}:\!u^{\mu}h^{\rho\sigma}_{\:,\mu}h^{\rho\sigma}_{\:,\nu}\!: 
-\frac{1}{4}:\!u^{\mu}_{,\mu}hh_{,\nu}\!:\\
&-\frac{1}{4}:\!u^{\mu}h_{,\mu}h_{,\nu}\!:
+\frac{1}{2}:\!u^{\mu}_{,\rho}h^{\mu\rho}h_{,\nu}\!:
-:\!u^{\mu}_{,\rho}h^{\rho\sigma}h^{\sigma\mu}_{\:,\nu}\!:+\ldots\Big\}\,.
\label{eq:d130}
\end{split}
\end{equation}
Then, $D_{2/1}^{\nu\:h}$ contains the following  terms that generate anomalies
\begin{multline}
D_{2/1}^{\nu\:h}(x,y)=\frac{\kappa^2}{4}\,\Big\{ 
+:\!u^{\mu}_{,\mu}h^{\rho\sigma}\f_{,\sigma}\f_{,\rho}\!: 
-\frac{m^2}{2} :\!u^{\mu}_{,\mu}h\f\f\!:
+:\!u^{\mu}h^{\rho\sigma}_{\:,\mu}\f_{,\sigma}\f_{,\rho}\!:+\\
-\frac{m^2}{2} :\!u^{\mu}h_{,\mu}\f\f\!:
-2 :\!u^{\mu}_{,\rho}h^{\rho\sigma}\f_{,\sigma}\f_{,\mu}\!: 
+ m^2 :\!u^{\mu}_{,\rho}h^{\mu\rho}\f\f\!:\Big\}\,\d^{\nu}_{x}\,D_{0}(x-y)\,.
\label{eq:d131}
\end{multline}
The first two fields in the normally ordered products depend on $x$, whereas the matter fields depend on $y$. The retarded part
$R_{2/1}^{\nu\:h}$ has exactly the same form as in~(\ref{eq:d131}) with the replacement $D_{0}\to D_{0}^{\text{ret}}$.
Applying $\d_{\nu}^{x}$ to $R_{2/1}^{\nu\:h}$~(\ref{eq:d131}), we obtain the local anomalies (local anomalies coming from 
$\d_{\nu}^{y}R_{2/2}^{\nu\:h}$ are just the same, therefore we get a factor two):
\begin{multline}
\mathrm{an}(\d_{\nu}^{x}R_{2/1}^{\nu\:h}+\d_{\nu}^{y}R_{2/2}^{\nu\:h})=\frac{\kappa^{2}}{2}\,\Big[
:\!u^{\mu}_{,\mu}h^{\rho\sigma}\f_{,\sigma}\f_{,\rho}\!: 
-\frac{m^2}{2} :\!u^{\mu}_{,\mu}h\f\f\!:
+:\!u^{\mu}h^{\rho\sigma}_{\:,\mu}\f_{,\sigma}\f_{,\rho}\!:+\\
-\frac{m^2}{2} :\!u^{\mu}h_{,\mu}\f\f\!:
- 2 :\!u^{\mu}_{,\rho}h^{\rho\sigma}\f_{,\sigma}\f_{,\mu}\!: 
+ m^2 :\!u^{\mu}_{,\rho}h^{\mu\rho}\f\f\!:\Big]\,\delta^{\sss (4)}(x-y)\,.
\label{eq:d132}
\end{multline}
Because of the $\delta$-function, all fields have now the same space-time dependence.

In the matter sector, the first term in the matter $Q$-vertex of Eq.~(\ref{eq:d21}) is the only one that can generate local anomalies, 
because it carries the $\nu$-index as a derivative. Then, $D_{2/1}^{\nu\:\sss M}$ becomes
\begin{equation}
D_{2/1}^{\nu\:\sss M}(x,y)=-\frac{\kappa^2}{2}\,\Big\{:\!u^{\mu}\f_{,\mu}h^{\rho\sigma}\f_{,\rho}\!:\,\d_{\sigma}^{x}\,\,
+\frac{m^2}{2} :\!u^{\mu}\f_{,\mu}h\f\!:\Big\}\,\d^{\nu}_{x} \,D_{m}(x-y)\,.
\label{eq:d133}
\end{equation}
The contribution coming from $D_{2/2}^{\nu\:\sss M}$ has $x\leftrightarrow y$ and the derivative attached to the first term is
$\d_{\sigma}^{y}$. The retarded part $R_{2/1}^{\nu\:\sss M}$ has the same form as in~(\ref{eq:d133}) with the replacement 
$D_{0} \to D_{0}^{\text{ret}}$. Since
\begin{multline}
:\!A(x)B(y)\!:\,\d_{\alpha}^{x}\delta^{\sss (4)}(x-y) +:\!A(y)B(x)\!:\,\d_{\alpha}^{y}\delta^{\sss (4)}(x-y)=\\
=-:\!A(x)_{,\alpha}B(x)\!:\,\delta^{\sss (4)}(x-y)+
:\!A(x)B(x)_{,\alpha}\!:\,\delta^{\sss (4)}(x-y)\,,
\label{eq:d134}
\end{multline}
by applying $\d^{\nu}_{x}$ to $R_{2/1}^{\nu\:\sss M}$ and $\d^{\nu}_{y}$ to $R_{2/2}^{\nu\:\sss M}$, we obtain
\begin{multline}
\mathrm{an}(\d_{\nu}^{x} R_{2/1}^{\nu\:\sss M}+\d_{\nu}^{y} R_{2/2}^{\nu\:\sss M})=\frac{\kappa^{2}}{2}\,\Big[
- m^2 :\!u^{\mu}h\f_{,\mu}\f\!: +:\!u^{\mu}_{,\sigma}h^{\rho\sigma}\f_{,\rho}\f_{,\mu} +\\
+:\!u^{\mu}h^{\rho\sigma}\f_{,\rho}\f_{,\mu\sigma}\!:
 - :\!u^{\mu}h^{\rho\sigma}_{\:,\sigma}\f_{,\rho}\f_{,\mu}\!:
- :\!u^{\mu}h^{\rho\sigma}\f_{,\rho\sigma}\f_{,\mu}\!:\Big]\,\delta^{\sss (4)}(x-y)\,.
\label{eq:d135}
\end{multline}
Because of the $\delta$-function, all fields have now the same space-time dependence and have been recast in the form $:\!uh\f\f\!:$.

\subsection{Quartic Normalization Terms}\label{sec:53}

Summing up, we have found all the local anomalies arising from $\d_{\nu}^{x} R_{2/1}^{\nu}+\d_{\nu}^{y} R_{2/2}^{\nu}$ attached to
normally ordered products of the type $:\!uh\f\f\!:$. These can be organized in three different types according to their Lorentz
structure apart from derivatives: type I $:\!u^{\mu}h^{\rho\sigma}\f\f\!:$ with derivatives $\d_{\mu}, \d_{\rho}, \d_{\sigma}$;
type II $:\!u^{\mu}h\f\f\!:$ with derivatives $\d_{\mu}, \d_{\rho}, \d_{\rho}$ and 
type III $:\!u^{\mu}h^{\mu\rho}\f\f\!:$ with derivatives $\d_{\rho}, \d_{\sigma}, \d_{\sigma}$. Different Lorentz types do not interfere.

From Eqs.~(\ref{eq:d132}) and~(\ref{eq:d135}), the local anomalies of type I are
\begin{multline}
\mathrm{an}(\d_{\nu}^{x} R_{2/1}^{\nu}+\d_{\nu}^{y} R_{2/2}^{\nu})\Big|_{\text{type I}}=\frac{\kappa^{2}}{2}\bigg[
+:\!u^{\mu}_{,\mu}h^{\rho\sigma}\f_{,\rho}\f_{,\sigma}\!:
+:\!u^{\mu}h^{\rho\sigma}_{\:,\mu}\f_{,\rho}\f_{,\sigma}\!:+\\
+:\!u^{\mu}h^{\rho\sigma}\f_{,\rho}\f_{,\mu\sigma}\!:
-:\!u^{\mu}_{,\rho}h^{\rho\sigma}\f_{,\rho}\f_{,\mu}\!:+\\
-:\!u^{\mu}h^{\rho\sigma}_{\:,\sigma}\f_{,\rho}\f_{,\mu}\!:
-:\!u^{\mu}h^{\rho\sigma}\f_{,\rho\sigma}\f_{,\mu}\!: \bigg]\,\delta^{\sss (4)}(x-y)\,.
\label{eq:d137}
\end{multline}
From Eqs.~(\ref{eq:d132}) and~(\ref{eq:d135}), the local anomalies of type II are
\begin{multline}
\mathrm{an}(\d_{\nu}^{x} R_{2/1}^{\nu}+\d_{\nu}^{y} R_{2/2}^{\nu})\Big|_{\text{type II}}=-\frac{\kappa^{2}}{2}\,\bigg[
\frac{m^2}{2} :\!u^{\mu}_{,\mu}h\f\f\!: +\frac{m^2}{2} :\!u^{\mu}h_{,\mu}\f\f\!:+\\ + m^2 :\!u^{\mu}h\f_{,\mu}\f\!:
\bigg]\,\delta^{\sss (4)}(x-y)\,,
\label{eq:d138}
\end{multline}
and those of type III are simply
\begin{equation}
\mathrm{an}(\d_{\nu}^{x} R_{2/1}^{\nu}+\d_{\nu}^{y} R_{2/2}^{\nu})\Big|_{\text{type III}}=\frac{\kappa^{2}}{2}\,\Big[
m^2 :\!u^{\mu}_{,\rho}h^{\mu\rho}\f\f\!:\Big]\,\delta^{\sss (4)}(x-y)\,.
\label{eq:d139}
\end{equation}

According to Eq.~(\ref{eq:d126}), the question is whether these local anomalies can be written as divergences and therefore
can be compensated by corresponding local divergence terms coming from $\d_{\nu}^{x}N_{2/1}^{\nu}+\d_{\nu}^{y}N_{2/2}^{\nu}$.
If it is not possible to reach such a compensation, normalization terms  on the left side of~(\ref{eq:d126}) have to be introduced
in order to preserve gauge invariance.

Due to the identity
\begin{multline}
\d_{\nu}^{x}\big(:\!A(x)B(y)\!:\,\delta^{\sss (4)}(x-y)\big) +\d_{\nu}^{y}\big(:\!A(y)B(x)\!:\,\delta^{\sss (4)}(x-y)\big)=\\
=:\!A(x)_{,\nu}B(x)\!:\,\delta^{\sss (4)}(x-y)  +  :\!A(x)B(x)_{,\nu}\!:\,\delta^{\sss (4)}(x-y)\,,
\label{eq:d140}
\end{multline}
the local anomalies of type I can be indeed written as a divergence:
\begin{multline}
\text{Eq.~(\ref{eq:d137})}\to
\d_{\mu}^{x}\big(:\!\underbrace{u^{\mu}h^{\rho\sigma}}_{x}\underbrace{\f_{,\rho}\f_{,\sigma}}_{y}\!:\,\delta^{\sss (4)}(x-y)\big)+
\d_{\mu}^{y}\big(:\!\underbrace{u^{\mu}h^{\rho\sigma}}_{y}\underbrace{\f_{,\rho}\f_{,\sigma}}_{x}\!:\,\delta^{\sss (4)}(x-y)\big)+\\
-\Big[ \d_{\rho}^{x}\big(:\!\underbrace{u^{\mu}h^{\rho\sigma}}_{x}\underbrace{\f_{,\mu}\f_{,\sigma}}_{y}\!:\,\delta^{\sss (4)}(x-y)\big)+
        \d_{\mu}^{y}\big(:\!\underbrace{u^{\mu}h^{\rho\sigma}}_{y}\underbrace{\f_{,\mu}\f_{,\sigma}}_{x}\!:\,\delta^{\sss (4)}(x-y)\big)\Big]\,.
\label{eq:d141}
\end{multline}
The same is true for the local anomalies of type II:
\begin{multline}
\text{Eq.~(\ref{eq:d138})}\to
m^2 \d_{\mu}^{x}\big(:\!\underbrace{u^{\mu}h}_{x}\underbrace{\f\f}_{y}\!:\,\delta^{\sss (4)}(x-y)\big)+
m^2 \d_{\mu}^{y}\big(:\!\underbrace{u^{\mu}h}_{y}\underbrace{\f\f}_{x}\!:\,\delta^{\sss (4)}(x-y)\big)\,.
\raisetag{17mm}\label{eq:d142}
\end{multline}
Therefore, by choosing the appropriate local divergence normalization terms $\d_{\nu}^{x}N_{2/1}^{\nu}+\d_{\nu}^{y}N_{2/2}^{\nu}$,
Eq.~(\ref{eq:d126}) restricted to the Lorentz type I and II holds.

The local anomaly of type III, Eq.~(\ref{eq:d139}) cannot be written as a divergence. Eq.~(\ref{eq:d126}) forces us to consider
normalization terms $N_{2}$. Then, we require that  the local anomaly of type III has to be coboundary, namely a gauge variation
of a normalization term $N_{2}$:
\begin{equation}
d_{Q}N_{2}(x,y)=\frac{\kappa^2 m^2}{2}\,:\!u^{\mu}(x)_{,\rho} h^{\mu\rho}(x)\f(x)\f(x)\!:\,\delta^{\sss (4)}(x-y)\,.
\label{eq:d143}
\end{equation}
The only possible $N_{2}$ that satisfies this requirement is
\begin{equation}
N_{2}(x,y)=\frac{i\kappa^2 m^2}{4}\,\Big\{ :\!h^{\alpha\beta}h^{\alpha\beta}\f\f\!: 
-\frac{1}{2}:\!hh\f\f\!:\Big\}\,\delta^{\sss (4)}(x-y)\,.
\label{eq:d144}
\end{equation}
Taking the factor $\frac{1}{2}$ for the second order $S$-matrix expansion~(\ref{eq:d1}) into account, the quartic interaction 
in~(\ref{eq:d144}), quadratic in $\kappa$, generated by gauge invariance agrees exactly with the term of order $\kappa^2$ in the
expansion of the matter Lagrangian density $\mathcal{L}_{\sss M}$ in~(\ref{eq:d26}). But in our case, this mechanism of
generation works in a purely quantum framework.

The only objection to this result is that this $N_{2}$ is not a `proper' normalization of a tree graph in $T_{2}$, obtained 
starting with $T_{1}^{h+u}$ and $T_{1}^{\sss M}$. But it can be considered as a normalization term of box-graphs in fourth order
with external legs $\sim :\!hh\f\f\!:$ constructed with two $T_{1}^{h+u}$ and two $T_{1}^{\sss M}$ or with four $T_{1}^{\sss M}$.

\subsection{Massless Matter Case}\label{sec:54}

For the massless matter coupling ${T}_{1}^{\sss M}(x)=i\frac{\kappa}{2} :\!h^{\alpha\beta}\f_{,\alpha}\f_{,\beta}\!:$,
perturbative gauge invariance to first order reads
\begin{equation}
d_{Q} T_{1}^{\sss M}(x)=\d_{\nu}^{x}\Big(\frac{\kappa}{2}\,\big\{:\!u^{\rho}\f_{,\rho}\f_{,\nu}\!:
-\frac{1}{2}:\!u^{\nu}\f_{,\rho}\f_{,\rho}\!:\big\}\Big)=:\d_{\nu}^{x}\,T_{1/1}^{\nu\; \sss M}(x)\,,
\label{eq:d147}
\end{equation}
namely, the matter $Q$-vertex of~(\ref{eq:d21}) in the $m\to 0$ limit. Performing the same calculation as for the massive case,
it turns out that the local anomalies are those of type I only, Eq.~(\ref{eq:d137}) which can be written as a divergence and therefore
compensated by proper normalization terms $\d_{\nu}^{x} N_{2/1}^{\nu}+\d_{\nu}^{y} N_{2/2}^{\nu}$ as in~(\ref{eq:d141}).
From Eq.~(\ref{eq:d126}), it follows that $d_{Q}N_{2}=0$, which is certainly  satisfied by $N_{2}=0$. This agrees with the fact that 
even classically there are no $h h \f \f$ couplings in the $m=0$ case: the expansion of $\mathcal{L}_{\sss M}$ written in terms of the 
Goldberg variable reads
\begin{equation}
\mathcal{L}_{\sss M}=\frac{1}{2}\f_{,\rho}\f^{,\rho}+\frac{\kappa}{2}\,h^{\mu\nu}\f_{,\mu}\f_{,\nu}\,.
\label{eq:d149}
\end{equation}
Therefore, $\mathcal{L}_{\sss M}^{\sss (2)} =0$. This concludes our discussion of the condition of perturbative gauge invariance 
to second order for tree graphs.

\section*{Acknowledgements}

I would like to thank Prof.~G.~Scharf, Adrian M\"uller and Mark Wellmann for valuable discussions and comments regarding these 
topics. 

\addcontentsline{toc}{section}{Appendix}

\addcontentsline{toc}{subsection}{Appendix 1: The \boldmath{$I_{m}^{\sss(\pm)}(p)_{\ldots}$}-Integrals}

\section*{Appendix 1: The \boldmath{$I_{m}^{\sss(\pm)}(p)_{\ldots}$}-Integrals}
\renewcommand{\theequation}{A.\arabic{equation}}
\setcounter{equation}{0}

The $I_{\sss m}^{\sss(\pm)}(p)_{\ldots}$-integrals are defined by
\begin{equation}
\begin{split}
I_{\sss m}^{\sss (\pm)}(p)_{- / \mu/\mu\nu/\mu\nu\rho/\mu\nu\rho\sigma}:=\int\!\!& d^{4}q \,\delta\big( (p-q)^2 -m^2 \big)\,
                    \Theta\big(\pm (p^{0}-q^{0})\big)\,\delta(q^{2} -m^2)\\
&\times\Theta(\pm q^{0})\,
\big[1,\,q_{\mu},\,q_{\mu}q_{\nu},\,q_{\mu}q_{\nu}q_{\rho},\,q_{\mu}q_{\nu}q_{\rho}q_{\sigma}\big]\,.
\label{eq:da1}
\end{split}
\end{equation}
Let us investigate  $I_{\sss m}^{\sss (+)}(p)$ in detail, because it contains the main feature of the calculation of the other
integrals. $p$ is time-like due to the presence of the $\Theta$- and $\delta$-distributions in the integrand. Therefore, we may choose
a Lorentz frame such that $p_{\nu}=(p_0 ,\bol{0})$, $p_0 >0$, then
\begin{equation}
\begin{split}
I_{\sss m}^{\sss (+)}(p_0)&=\int\!\!d^{4}q\,\delta(p_0^2-2p_0q_0+q_0^2-|\bol{q}|^2 -m^2)\,\Theta(p_0-q_0)\delta(q^2-m^2)\,\Theta(+q_0)\\
&=\int\!\!d^{3}q\,\frac{\delta(p_0^2-2p_0 E_{\bol{q}})\,\Theta(p_0-E_{\bol{q}})}{2E_{\bol{q}}}\,,
\label{eq:da2}
\end{split}
\end{equation}
with $E_{\bol{q}}=q_0=\sqrt{\bol{q}^2 +m^2}$. Then
\begin{equation}
\begin{split}
I_{\sss m}^{\sss (+)}(p_0)&=4\pi\,\int_{0}^{\infty}\!\!d|\bol{q}|\,\frac{|\bol{q}|^2}{2E_{\bol{q}}}\,\frac{1}{2p_0}\,
\delta\left(\frac{p_0}{2}-E_{\bol{q}}\right)\,\Theta(p_0-E_{\bol{q}})\\
&=\frac{\pi}{p_0}\,\Theta(p_0^2 -4m^2)\,\Theta(p_0)\,\int_{0}^{\infty}\!\!d|\bol{q}|\,\frac{|\bol{q}|^2}{E_{\bol{q}}}\,
\delta\left(\frac{p_0}{2}-E_{\bol{q}}\right)\,,
\label{eq:da3}
\end{split}
\end{equation}
because of the $\delta$-distribution
\begin{equation}
E_{\bol{q}}=\sqrt{\bol{q}^2 +m^2}=\frac{p_0}{2}\Longrightarrow |\bol{q}|=\sqrt{\frac{p_0^2}{4}-m^2}\,.
\label{eq:da4}
\end{equation}
Therefore
\begin{equation}
\begin{split}
I_{\sss m}^{\sss (+)}(p_0)&=\frac{\pi}{p_0}\,\Theta(p_0^2 -4m^2)\,\Theta(p_0)\,\sqrt{\frac{p_0^2}{4}-m^2}\\
&=\frac{\pi}{2}\,\sqrt{1-\frac{4m^2}{p_0^2}}\,\Theta(p_0^2 -4m^2)\,\Theta(p_0)\,.
\label{eq:da5}
\end{split}
\end{equation}
$I_{\sss m}^{\sss (-)}$ can be calculated analogously and the result in an arbitrary Lorentz frame reads
\begin{equation}
I_{\sss m}^{\sss (\pm)}(p)=\frac{\pi}{2}\,\sqrt{1-\frac{4m^2}{p^2}}\,\Theta(p^2 -4m^2)\,\Theta(\pm p_0)=:\frac{\pi}{2}\,
\hat{d}(p)^{\sss (\pm)}_{\sss m} \,.
\label{eq:da6}
\end{equation}

Computing $I_{\sss m}^{\sss (\pm)}(p)_{\mu}$ for $p_{\nu}=(p_0 ,\bol{0})$, $p_0 >0$, we have a non-vanishing contribution only for 
$\mu =0$. An additional factor $q_0$ is therefore present in the integrand in~(\ref{eq:da3}), which is set equal to $E_{\bol{q}}$ and then
to $\frac{p_0}{2}$. This leads to
\begin{equation}
I_{\sss m}^{\sss (\pm)}(p)_{\mu}=p_{\mu}\,\frac{\pi}{4}\,\sqrt{1-\frac{4m^2}{p^2}}\,\Theta(p^2 -4m^2)\,\Theta(\pm p_0)=\frac{\pi}{4}\,
p_{\mu}\,\hat{d}(p)^{\sss (\pm)}_{\sss m}\,,
\label{eq:da7}
\end{equation}
in an arbitrary Lorentz frame.

For $I_{\sss m}^{\sss (\pm)}(p)_{\mu\nu}$, we have to take the two conditions
\begin{equation}
\begin{cases}
&I_{\sss m}^{\sss (\pm)}(p)_{\mu}^{\ \mu}=m^2\,I_{\sss m}^{\sss (\pm)}(p)\,,\\
&p^{\mu}p^{\nu}\,I_{\sss m}^{\sss (\pm)}(p)_{\mu\nu}=\frac{p^4}{4}\,I_{\sss m}^{\sss (\pm)}(p)\,,
\label{eq:da8}
\end{cases}
\end{equation}
into account. The first condition is imposed by the presence of $\delta(q^2-m^2)$ in~(\ref{eq:da1}). Computing 
$p^{\mu}p^{\nu}\,I_{\sss m}^{\sss (+)}(p)_{\mu\nu}$ for $p_{\nu}=(p_0 ,\bol{0})$, $p_0 >0$, we get an additional factor
$(p_0 E_{\bol{q}})^2$ in the integrand~(\ref{eq:da3}) which gives $\frac{p_0^4}{4}$ due to $\delta(\frac{p_0}{2}-E_{\bol{q}})$. In
a general Lorentz frame, the second condition in~(\ref{eq:da8}) must hold. Therefore, making the ansatz
\begin{equation}
I_{\sss m}^{\sss (\pm)}(p)_{\mu\nu}=\big( a(p^2)\,p_{\mu}p_{\nu}+b(p^2)\,p^2\,\eta_{\mu\nu}\big)\,I_{\sss m}^{\sss (\pm)}(p)\,,
\label{eq:da9}
\end{equation}
the conditions~(\ref{eq:da8}) imply
\begin{equation}
a(p^2)=\frac{1}{3}-\frac{m^2}{3p^2}\,,\qquad   b(p^2)=\frac{-1}{12}+\frac{m^2}{3p^2}\,,
\label{eq:da10}
\end{equation}
so that
\begin{equation}
I_{\sss m}^{\sss (\pm)}(p)_{\mu\nu}=\Bigg\{\left(1-\frac{m^2}{p^2}\right)\,p_{\mu}p_{\nu}-
\left(\frac{1}{4}-\frac{m^2}{p^2}\right)\,p^2\,\eta_{\mu\nu}\Bigg\}\,\frac{\pi}{6}\,\hat{d}(p)^{\sss (\pm)}_{\sss m}\,.
\label{eq:da11}
\end{equation}

For $I_{\sss m}^{\sss (\pm)}(p)_{\mu\nu\rho}$, we have to take the two conditions
\begin{equation}
\begin{cases}
&I_{\sss m}^{\sss (\pm)}(p)_{\mu\nu}^{\quad \nu}=m^2\,I_{\sss m}^{\sss (\pm)}(p)_{\mu}\,,\\
&p^{\mu}p^{\nu}p^{\rho}\,I_{\sss m}^{\sss (\pm)}(p)_{\mu\nu\rho}=\frac{p^6}{8}\,I_{\sss m}^{\sss (\pm)}(p)\,,
\label{eq:da12}
\end{cases}
\end{equation}
into account. The first condition follows from the definitions of the $I_{\sss m}^{\sss (\pm)}$-integrals. Computing
$p^{\mu}p^{\nu}p^{\rho}\,I_{\sss m}^{\sss (+)}(p)_{\mu\nu\rho}$ for $p_{\nu}=(p_0 ,\bol{0})$, $p_0 >0$, we obtain an additional factor
$(p_0 E_{\bol{q}})^3$ in the integrand~(\ref{eq:da3}) which yields  $\frac{p_0^6}{8}$. Making the ansatz
\begin{equation}
I_{\sss m}^{\sss (\pm)}(p)_{\mu\nu\rho}=\Big( c(p^2)\,p_{\mu}p_{\nu}p_{\rho}+d(p^2)\,p^2\,\big(p_{\rho}\eta_{\mu\nu}
+p_{\mu}\eta_{\rho\nu}+p_{\nu}\eta_{\rho\mu}\big)\Big)\,I_{\sss m}^{\sss (\pm)}(p)\,,
\label{eq:da13}
\end{equation}
then~(\ref{eq:da12}) imply
\begin{equation}
c(p^2)=\frac{1}{4}-\frac{m^2}{2p^2}\,,\qquad   c(p^2)=\frac{-1}{24}+\frac{m^2}{6p^2}\,,
\label{eq:da14}
\end{equation}
and therefore
\begin{equation}
\begin{split}
I_{\sss m}^{\sss (\pm)}(p)_{\mu\nu\rho}&=\Bigg\{\left(1-\frac{2m^2}{p^2}\right)\,p_{\mu}p_{\nu}p_{\rho}-
\left(\frac{1}{6}-\frac{2m^2}{3p^2}\right)\,p^2\,\big(p_{\rho}\eta_{\mu\nu}
+p_{\mu}\eta_{\rho\nu}+p_{\nu}\eta_{\rho\mu}\big)\Bigg\}\\
&\qquad\qquad \times\frac{\pi}{8}\,\hat{d}(p)^{\sss (\pm)}_{\sss m}\,.
\label{eq:da15}
\end{split}
\end{equation}

For $I_{\sss m}^{\sss (\pm)}(p)_{\mu\nu\rho\sigma}$, we repeat this calculational scheme again. For the 
reasons pointed out above, now three conditions
\begin{equation}
\begin{cases}
&I_{\sss m}^{\sss (\pm)}(p)_{\mu\nu\rho}^{\quad\  \rho}=m^2\,I_{\sss m}^{\sss (\pm)}(p)_{\mu\nu}\,,\\
&I_{\sss m}^{\sss (\pm)}(p)_{\mu\ \rho}^{\ \mu\  \rho}=m^4\,I_{\sss m}^{\sss (\pm)}(p)\,,\\
&p^{\mu}p^{\nu}p^{\rho}p^{\sigma}\,I_{\sss m}^{\sss (\pm)}(p)_{\mu\nu\rho\sigma}=\frac{p^8}{16}\,I_{\sss m}^{\sss (\pm)}(p)\,,
\label{eq:da16}
\end{cases}
\end{equation}
must hold. Therefore, the ansatz
\begin{multline}
I_{\sss m}^{\sss (\pm)}(p)_{\mu\nu\rho\sigma}=\Big( e(p^2)\,p_{\mu}p_{\nu}p_{\rho}p_{\sigma}+f(p^2)\,p^2\,
\big(p_{\rho}p_{\sigma}\eta_{\mu\nu}+p_{\rho}p_{\mu}\eta_{\sigma\nu}+p_{\rho}p_{\nu}\eta_{\sigma\mu}+\\+p_{\mu}p_{\sigma}\eta_{\rho\nu}+
p_{\nu}p_{\sigma}\eta_{\rho\mu}+p_{\mu}p_{\nu}\eta_{\rho\sigma}\big)+g(p^2)\,p^4\,\big(\eta_{\mu\rho}\eta_{\nu\sigma}+
\eta_{\mu\sigma}\eta_{\nu\rho}+\eta_{\mu\nu}\eta_{\rho\sigma}\big)\Big)\,I_{\sss m}^{\sss (\pm)}(p)\,,
\label{eq:da17}
\end{multline}
leads to 
\begin{gather}
e(p^2)=\frac{1}{5}-\frac{3m^2}{5p^2}+\frac{m^4}{5p^4}\,,\qquad f(p^2)=\frac{-1}{40}+\frac{7m^2}{60p^2}-\frac{m^4}{15p^4}\,,\nonumber\\
g(p^2)=\frac{1}{15}\left(\frac{1}{16}- \frac{m^2}{2p^2}+\frac{m^4}{p^4}\right)\,,
\label{eq:da18}
\end{gather}
so that 
\begin{equation}
\begin{split}
I_{\sss m}^{\sss (\pm)}(p)_{\mu\nu\rho\sigma}&=\Bigg\{\left(1-\frac{3m^2}{p^2}+\frac{m^4}{p^4}\right) \,p_{\mu}p_{\nu}p_{\rho}p_{\sigma}
-\left(\frac{1}{8}-\frac{7m^2}{12p^2}+\frac{m^4}{3p^4} \right)\, p^2\\
&\times \big(p_{\rho}p_{\sigma}\eta_{\mu\nu}+p_{\rho}p_{\mu}\eta_{\sigma\nu}+p_{\rho}p_{\nu}\eta_{\sigma\mu}
+p_{\mu}p_{\sigma}\eta_{\rho\nu}+
p_{\nu}p_{\sigma}\eta_{\rho\mu}+p_{\mu}p_{\nu}\eta_{\rho\sigma}\big)\\
&+\left(\frac{1}{48}- \frac{m^2}{6p^2}+\frac{m^4}{3p^4}\right)\,p^4\,
\big(\eta_{\mu\rho}\eta_{\nu\sigma}+\eta_{\mu\sigma}\eta_{\nu\rho}+\eta_{\mu\nu}\eta_{\rho\sigma}\Big)\Bigg\}\,\frac{\pi}{10}\,
\hat{d}(p)^{\sss (\pm)}_{\sss m}\,.
\label{eq:da19}
\end{split}
\end{equation}

\addcontentsline{toc}{subsection}{Appendix 2: The \boldmath{$\tilde{I}^{\sss(\pm)}(p)_{\ldots}$}-Integrals}

\section*{Appendix 2: The \boldmath{$\tilde{I}^{\sss(\pm)}(p)_{\ldots}$}-Integrals}
\renewcommand{\theequation}{B.\arabic{equation}}
\setcounter{equation}{0}

The products of Jordan--Pauli distributions of Eq.~(\ref{eq:d98}) are evaluated in momentum space:
\begin{equation}
\begin{split}
\hat{C}_{\cdot|\cdot}^{\sss (\pm)}(p)&=\frac{1}{(2\pi)^2}\,\int\!\!d^{4}q\,\hat{D}_{0}^{\sss (\pm)}(p-q)\,\hat{D}_{m}^{\sss (\pm)}(q)\\
&=\frac{-1}{(2\pi)^4}\,\int\!\!d^{4}q\,\delta\big((p-q)^2\big)\,\Theta\big(\pm(p_0-q_0)\big)\,\delta(q^2-m^2)\,\Theta(\pm q_0)\\
&=:\frac{-1}{(2\pi)^4}\,\tilde{I}^{\sss (\pm)}(p)\,,
\label{eq:db1}
\end{split}
\end{equation}
and 
\begin{equation}
\begin{split}
\hat{C}_{\cdot|\alpha}^{\sss (\pm)}(p)&=\frac{1}{(2\pi)^2}\,\int\!\!d^{4}q\,\hat{D}_{0}^{\sss (\pm)}(p-q)\,\big(-i\,q_{\alpha}\big)\,
     \hat{D}_{m}^{\sss (\pm)}(q)\\
&=\frac{+i}{(2\pi)^4}\,\int\!\!d^{4}q\,q_{\alpha}\,\delta\big((p-q)^2\big)\,\Theta\big(\pm(p_0-q_0)\big)\,\delta(q^2-m^2)\,
\Theta(\pm q_0)\\
&=:\frac{+i}{(2\pi)^4}\,\tilde{I}^{\sss (\pm)}(p)_{\alpha}\,.
\label{eq:db2}
\end{split}
\end{equation}
Let us calculate $\tilde{I}^{\sss (+)}(p)$. From the $\delta$- and $\Theta$-distributions, it follows that $p$ is time-like. We choose
a Lorentz frame in which $p_{\nu}=(p_0,\bol{0})$, $p_0 >0$. Then
\begin{equation}
\begin{split}
\tilde{I}^{\sss (+)}(p_0)&=\int\!\!d^{4}q\,\delta(p_0^2-2p_0 q_0 +q_0^2-\bol{q}^2)\,\Theta(p_0-q_0)\,\Theta(q_0)
\frac{\delta(q_{0}-E_{\bol{q}}) }{2E_{\bol{q}}}\\
&=\int\!\!\frac{d^3 q}{2E_{\bol{q}}}\,\delta(p_0^2 -2p_0 E_{\bol{q}} +m^2)\,\Theta(p_0 -E_{\bol{q}})\,.
\label{eq:db3}
\end{split}
\end{equation}
With $E_{\bol{q}}=\sqrt{\bol{q}^2 +m^2}$. The $\delta$-distribution implies $p_0=E_{\bol{q}} +|\bol{q}|$. 
From $E_{\bol{q}}^2=\bol{q}^2 +m^2=(p_0 -|\bol{q}|)^2$, we obtain $|\bol{q}|=(p_0^2 -m^2)/2p_0$, which, due to $p_0>0$, yields
$\Theta(p_0^2-m^2)$ and
\begin{equation}
\begin{split}
\tilde{I}^{\sss (+)}(p_0)&=2\pi\,\Theta(p_0^2 -m^2)\,\Theta(p_0)\,\int_{0}^{\infty}\!\!d|\bol{q}|\,\frac{|\bol{q}|^2}{E_{\bol{q}}}\,
             \delta(p_0^2 -2p_0 E_{\bol{q}} +m^2)\\
&=2\pi\,\Theta(p_0^2 -m^2)\,\Theta(p_0)\,\int_{m}^{\infty}\!\!dE_{\bol{q}}\,\frac{|\bol{q}|}{2p_0}\,
   \delta(E_{\bol{q}}-\frac{p_0^2+m^2}{2p_0})\\
&=\frac{\pi}{p_0}\,\Theta(p_0^2 -m^2)\,\Theta(p_0)\,\sqrt{E_{\bol{q}}^2 -m^2}\Big|_{E_{\bol{q}}=\frac{p_0^2+m^2}{2p_0}}\\
&=\frac{\pi}{2p_0^2}\,\Theta(p_0^2 -m^2)\,\Theta(p_0)\,(p_0^2 -m^2)=
    \frac{\pi}{2}\,\Theta(p_0^2 -m^2)\,\Theta(p_0)\,(1 -\frac{m^2}{p_0^2})\,.
\raisetag{2cm}\label{eq:db6}
\end{split}
\end{equation}
Therefore, in an arbitrary Lorentz frame:
\begin{equation}
\tilde{I}^{\sss (\pm)}(p)=\frac{\pi}{2}\,\Theta(p^2 -m^2)\,\Theta(\pm p_0)\,\left(1 -\frac{m^2}{p^2}\right)\,.
\label{eq:db7}
\end{equation}
The second integral can be in a similar manner computed: in the Lorentz frame with $p_{\nu}=(p_0,\bol{0})$, $p_0 >0$, 
$\tilde{I}^{\sss (+)}(p_0)_{i}$, $i=1,2,3$, vanishes for  symmetry reasons, then
\begin{equation}
\begin{split}
\tilde{I}^{\sss (+)}(p_0)_{0}&=\int\!\!\frac{d^3 q}{2E_{\bol{q}}}\,E_{\bol{q}}\,\delta(p_0^2 -2p_0 E_{\bol{q}} +m^2)\,
    \Theta(p_0 -E_{\bol{q}})\\
&=2\pi\,\Theta(p_0^2 -m^2)\,\Theta(p_0)\,\int_{0}^{\infty}\!\!d|\bol{q}|\,\frac{|\bol{q}|^2}{2p_0 E_{\bol{q}}}\,E_{\bol{q}}\,
             \delta(E_{\bol{q}}-\frac{p_0^2+m^2}{2p_0} )\\
&=\frac{\pi}{p_0}\,\Theta(p_0^2 -m^2)\,\Theta(p_0)\,\int_{m}^{\infty}\!\!dE_{\bol{q}}\,|\bol{q}|\,E_{\bol{q}}\,
   \delta(E_{\bol{q}}-\frac{p_0^2+m^2}{2p_0} )\\
&=\frac{\pi}{p_0}\,\Theta(p_0^2 -m^2)\,\Theta(p_0)\,\frac{p_0^2 -m^2}{2p_0}\,\frac{p_0^2+m^2}{2p_0}\\
&=\frac{\pi}{4}\,p_0\,\Theta(p_0^2 -m^2)\,\Theta(p_0)\,\left( 1-\frac{m^2}{p_0^2}\right)\,\left( 1+\frac{m^2}{p_0^2}\right)\,.
\label{eq:db8}
\end{split}
\end{equation}
Therefore, in an arbitrary Lorentz frame we have
\begin{equation}
\tilde{I}^{\sss (\pm)}(p)_{\alpha}=\frac{\pi}{4}\,p_{\alpha}\,\left(1 -\frac{m^4}{p^4}\right)\,\Theta(p^2 -m^2)\,\Theta(\pm p_0)\,.
\label{eq:db9}
\end{equation}

\end{document}